%% file: burst-deletions_Journal_Final_Single_Col.tex
\documentclass[letter,onecolumn]{IEEEtran}
\usepackage[utf8]{inputenc}
\usepackage{graphicx}
\usepackage{amssymb}
\usepackage[cmex10]{amsmath}
\usepackage{color}
\usepackage{theorem}
\usepackage{hyperref}
\usepackage{url}
\usepackage{breakurl}
\usepackage{booktabs}
\usepackage{bbm}
\usepackage{bbold}
\usepackage{comment}

\usepackage[english]{babel}
\usepackage[english = american]{csquotes}
\usepackage{mathtools}
\usepackage{mathdots}
\MakeOuterQuote{"} 

\definecolor{light_gray}{rgb}{0.6,0.6,0.6}
\definecolor{awgray}{rgb}{0.7,0.7,0.7}
\definecolor{awgray_dark}{rgb} {0.4,0.4,0.4}

\usepackage{pgf}
\usepackage{tikz}
\usetikzlibrary{matrix,chains,positioning,arrows,calc,decorations,plotmarks,patterns, fit,backgrounds}
\usepackage{pgfplots}
\usepgfplotslibrary{external}
\usetikzlibrary{external}                       
\tikzsetexternalprefix{tikzpic/}                   

\pgfplotscreateplotcyclelist{mylist}{red,blue,black,yellow,brown}

\tikzset{
    >=stealth',
    mycircle/.style={circle, draw=black, thick, text width=.1em, minimum height=.8em, text centered, font=\scriptsize}, 
    mycircle_gray/.style={circle, draw=gray, thick, text width=.1em, minimum height=.8em, text centered, font=\tiny},    
    mycircle_small/.style={circle,draw=awgray_dark,fill = awgray_dark, inner sep=0,minimum size=.6em},       
    mycircle_small_black/.style={circle,draw=black,fill = black, inner sep=0,minimum size=.6em},   
    mybox/.style={rectangle,rounded corners,draw=black, thick,text width=1em,minimum height=4em,minimum width=4em,text centered},     
    mybox_small/.style={rectangle,rounded corners,draw=black, thick,text width=1em,minimum height=2em,minimum width=2em,text centered},               
    mybox_vec/.style={rectangle,rounded corners,draw=black, thick,text width=1em,minimum height=0.7em, minimum width=4em,text centered},  
    mybox_vec_short/.style={rectangle,rounded corners,draw=black, thick,text width=1em,minimum height=0.7em, minimum width=2em,text centered},                  
    pil/.style={->, thick, shorten <=2pt, shorten >=2pt,},
}

\usepackage{algorithm,algpseudocode}

\makeatletter
\newcommand{\removelatexerror}{\let\@latex@error\@gobble}
\makeatother

\IEEEoverridecommandlockouts

\input{mydefs.tex}

\begin{document}

\title{Codes Correcting a Burst of Deletions or Insertions}
\author{\IEEEauthorblockN{Clayton Schoeny, Antonia Wachter-Zeh, Ryan Gabrys, and Eitan Yaakobi}\\
\thanks{Part of the results in the paper will be presented at the IEEE International Symposium on Information Theory, July 2016 \cite{BurstDeletions-ISIT}.
C. Schoeny's work was funded in part by the NISE program at SSC Pacific. A. Wachter-Zeh was supported by the European Union's Horizon 2020 research and innovation programme under the Marie Sklodowska-Curie grant agreement No. 655109. R. Gabrys' work was funded in part by the NISE program at SSC Pacific. E. Yaakobi's work was supported in part by the Israel Science Foundation (ISF) grant No. 1624/14.}
\thanks{C. Schoeny is with the Department of Electrical Engineering, University of California, Los Angeles, CA 90095 USA (email: cschoeny@ucla.edul).}
\thanks{R. Gabrys is with Spawar Systems Center, San Diego, CA 92152 USA (e-mail: ryan.gabrys@navy.mil).}
\thanks{A. Wachter-Zeh and E. Yaakobi are with the Computer Science Department, Technion--Israel Institute of Technology, Haifa 32000, Israel
(e-mails: antonia@cs.technion.ac.il, yaakobi@cs.technion.ac.il).}
}
\maketitle

\begin{abstract}
This paper studies codes that correct bursts of deletions. Namely, a code will be called a \emph{$b$-burst-deletion-correcting code} if it can correct a deletion of any $b$ consecutive bits. While the lower bound on the redundancy of such codes was shown by Levenshtein to be asymptotically $\log(n)+b-1$, the redundancy of the best code construction by Cheng \textit{et al.} is $b(\log (n/b+1))$. In this paper we close on this gap and provide codes with redundancy at most $\log(n) + (b-1)\log(\log(n)) +b -\log(b)$.

We also derive a non-asymptotic upper bound on the size of $b$-burst-deletion-correcting codes and extend the burst deletion model to two more cases: 1) A deletion burst of at most $b$ consecutive bits and 2) A deletion burst of size at most $b$ (not necessarily consecutive). We extend our code construction for the first case and study the second case for $b=3,4$. The equivalent models for insertions are also studied and are shown to be equivalent to correcting the corresponding burst of deletions.
\end{abstract}


\section{Introduction}\label{sec:introduction}
In communication and storage systems, symbols are often inserted or deleted due to synchronization errors. These errors can be caused by a variety of disturbances such as timing defects or packet-loss. Constructing codes that correct insertions or deletions is a notoriously challenging problem since a relatively small number of edits can cause the transmitted and received sequences to be vastly different in terms of the Hamming metric.

For disconnected, intermittent, and low-bandwidth environments, the problem of recovering from symbol insertion/deletion errors becomes exacerbated \cite{dandashi2007tactical}. From the perspective of the communication systems, these errors manifest themselves in bursts where the errors tend to cluster together. Our goal in this work is the study of codes capable of correcting bursts of insertion/deletion errors. Such codes have many applications pertaining to the synchronization of data in wireless sensor networks and satellite communication devices \cite{jeong2003forward}.

In the 1960s, Varshamov, Tenengolts, and Levenshtein laid the foundations for codes capable of correcting insertions and deletions. In 1965, Varshamov and Tenengolts created a class of codes (now known as VT-codes) that is capable of correcting asymmetric errors on the Z-channel \cite{tenengolts1984nonbinary,VarshTene-SingleDeletion1965}. Shortly thereafter, Levenshtein proved that these codes can also be used to correct a single insertion or deletion \cite{Levenshtein-binarycodesCorrectingDeletions} and he also constructed a class of codes that can correct two adjacent insertions or deletions \cite{Levenshtein-TwoAdjacentDeletions}.

The main goal of this work is to study codes that correct a \emph{burst of deletions} which refers to the deletion of a fixed number of consecutive bits. A code will be called a \emph{$b$-burst-deletion-correcting code} if it can correct any deletion burst of size $b$. For example, the codes studied by Levenshtein in~\cite{Levenshtein-TwoAdjacentDeletions} are two-burst-deletion-correcting codes. 

Establishing tight upper bounds on the cardinality of burst-deletion-correcting codes is a challenging task since the burst deletion balls are not all of the same size. In \cite{Levenshtein-binarycodesCorrectingDeletions}, Levenshtein derived an asymptotic upper bound on the maximal cardinality of a $b$-burst-deletion-correcting code, given by $\frac{2^{n-b+1}}{n}$. Therefore, the minimum redundancy of such a code should be approximately $\log(n)+b-1$. Using the method developed recently by Kulkarni and Kiyavash in \cite{KulkarniKiyavash-UpperBoundsForDeletion} for deriving an upper bound on deletion-correcting codes, we establish a non-asymptotic upper bound on the cardinality of $b$-burst-deletion-correcting codes which matches the asymptotic upper bound by Levenshtein. 

On the other hand, the best construction of $b$-burst-deletion-correcting codes, that we are aware of, is Construction 1 by Cheng \textit{et al}.~\cite{ChengSwartFerreiraAbdelGhaffar-ThreeOrMoreBurstDelIns}. The redundancy of this construction is $b(\log (n/b+1))$ and therefore there is still a significant gap between the lower bound on the redundancy and the redundancy of this construction. One of our main results in this paper is showing how to improve the construction from~\cite{ChengSwartFerreiraAbdelGhaffar-ThreeOrMoreBurstDelIns} and deriving codes whose redundancy is at most
\begin{equation}\label{eq:red_result}
	\log(n) + (b-1)\log(\log(n)) +b -\log(b),
\end{equation}
which is larger than the lower bound on the redundancy by roughly $(b-1)\log(\log(n))$.

This paper is organized as follows. In Section \ref{sec:preliminaries}, we define the common terms used throughout the paper and we detail the previous results that will be used as a comparison. In particular, we present two additional models: 1) A deletion burst of at most $b$ consecutive bits and 2) A non-consecutive deletion burst of size at most $b$. We also extend these definitions to insertions. Then, in Section~\ref{sec:equivalence}, we prove the equivalence between correcting insertions and deletions in each of the three burst models studied in the paper. We dedicate Section \ref{sec:bound} to deriving an explicit upper bound on the code cardinality of $b$-burst-deletion-correcting codes using techniques developed by Kulkarni and Kiyavash~\cite{KulkarniKiyavash-UpperBoundsForDeletion}. Note that in the asymptotic regime, our bound yields the bound established by Levenshtein \cite{Levenshtein-binarycodesCorrectingDeletions}. In Section \ref{sec:exactly_b}, we construct $b$-burst-deletion-correcting codes with the redundancy stated in~(\ref{eq:red_result}). In Sections \ref{sec:at_most_b} and~\ref{sec:non_consecutive}, we present code constructions that correct a deletion burst of size at most $ b $ and codes that correct a non-consecutive burst of size at most three and four, respectively. Lastly, Section~\ref{sec:conc} concludes the paper and lists some open problems in this area.

\section{Preliminaries and Previous Work}\label{sec:preliminaries}
\subsection{Notations and Definitions}
Let $\Fq$ be a finite field of order $q$, where $q$ is a power of a prime and let $\Fq^n$ denote the set of all vectors (sequences) of length $n$ over $\Fq$.
Throughout this paper, we restrict ourselves to binary vectors, i.e., $q=2$.
A \emph{subsequence} of a vector $\vec{x} = (x_1,x_2,\dots,x_n)$ is formed by taking a subset of the symbols of $\vec{x}$ and aligning them without changing their order. Hence, any vector $\vec{y} = (x_{i_1},x_{i_2}, \dots, x_{i_{m}} )$ is a subsequence of $\vec{x}$ if $1 \leq i_1 < i_2 < \dots < i_m \leq n$, and in this case we say that $n-m$ \emph{deletions} occurred in the vector $\vec{x}$ and $\vec{y}$ is the result.

A \emph{run} of length $r$ of a sequence $\vec{x}$ is a subvector of $\vec{x}$ such that $x_i=x_{i+1}=\dots=x_{i+r-1}$,
in which $x_{i-1} \neq x_i$ if $i>1$, and if $i+r-1 <n$, then $x_{i+r-1} \neq x_{i+r}$. We denote by $r(\vec{x})$ the number of runs of a sequence $\vec{x} \in \FTwo^n$. 

We refer to a \emph{deletion burst of size $b$} when exactly $b$ consecutive deletions have occurred, i.e., from $\vec{x}$, we obtain a subsequence $(x_1,\dots,x_{i},x_{i+b+1},\dots,x_n) \in \FTwo^{n-b}$. Similarly, a \emph{deletion burst of size at most $b$} results in a subsequence $(x_1,\dots,x_{i},x_{i+a+1},\dots,x_n) \in \FTwo^{n-a}$, for some $a \leq b$.
More generally, a \emph{non-consecutive deletion burst of size at most $b$} is the event where within $b$ consecutive symbols of $\vec{x}$, there were some $a \leq b$ deletions, i.e., we obtain a subsequence $(x_1,\dots,x_{i},x_{i+i_1},x_{i+i_2}, \dots,x_{i+i_{b-a}},x_{i+b+1},\dots,x_n) \in \FTwo^{n-a}$, for some $a \leq b$, where $1 \leq i_1 < i_2 < \dots < i_{b-a}\leq b$.

The \emph{$b$-burst-deletion ball} of a vector $\vec{x} \in \FTwo^n$, is denoted by $D_b(\vec{x})$, and is defined to be the set of subsequences of $\vec{x}$ of length $n-b$ obtained by the deletion of a burst of size $b$. Similarly, $D_{\leq b}(\vec{x})$ is defined to be the set of subsequences of $\vec{x}$ obtained from a deletion burst of size at most $b$. 

A \emph{$b$-burst-deletion-correcting code} $\mycode{C}$ is a set of codewords in $\FTwo^n$ such that there are no two codewords in $\mycode{C}$ where deletion bursts of size $b$ result in the same word of length $n-b$. That is, for every $\vec{x},\vec{y}\in \mycode{C}$, $D_b(\vec{x})\cap D_b(\vec{y}) = \emptyset$.

We will use the following notations for bursts of insertions, namely: \emph{insertions burst of size (at most) $b$}, \emph{$b$-burst-insertion ball}, and \emph{$b$-burst-insertion-correcting code}.

Throughout this paper, we let $b$ be a fixed integer which divides $n$. Similar to \cite{ChengSwartFerreiraAbdelGhaffar-ThreeOrMoreBurstDelIns}, for a vector $\vec{x} = (x_1,x_2,\dots,x_n)$, we define the following~$b \times \frac{n}{b}$ array:
\begin{equation*}
A_b(\vec{x}) = 
\begin{bmatrix}
x_1 & x_{b+1} &  \dots  & x_{n-b+1} \\
x_2 & x_{b+2} &  \dots  & x_{n-b+2} \\
\vdots & \vdots &  \ddots & \vdots \\
x_b & x_{2b} &  \dots  & x_n
\end{bmatrix},
\end{equation*}
and for $1\leq i\leq b$ we denote by $A_b(\vec{x})_i$ the $i$th row of the array $A_b(\vec{x})$.

For two vectors $\vec{x}$, $\vec{y} \in \FTwo^n$, the \emph{Levenshtein distance} $d_L(\vec{x},\vec{y})$ is the minimum number of insertions and deletions that is necessary to change $\vec{x}$ into $\vec{y}$. Unless stated otherwise, all logarithms in this paper are taken according to base 2.
\subsection{Previous Work}\label{subsec:rev_work}
In this subsection, we recall known results on codes which correct deletions and insertions. These results will be used later as a comparison reference for our constructions.
\subsubsection{Single-deletion-correcting codes}
The Varshamov-Tenengolts (VT) codes \cite{VarshTene-SingleDeletion1965} are a family of single-deletion-correcting codes (see also Sloane's survey in~\cite{Sloane01onsingle-deletion-correcting}) and are defined as follows.
\begin{definition}
For $0 \leq a \leq n$, the Varshamov-Tenengolts (VT) code $VT_{a}(n)$ is defined to be the following set of binary vectors:
\begin{equation*}
VT_{a}(n) \triangleq \bigg\{\vec{x}=(x_1,\ldots,x_n) \ : \ \sum_{i=1}^{n}i x_i \equiv a~(\bmod (n+1))\bigg\}.
\end{equation*}
\end{definition}
Levenshtein proved in~\cite{Levenshtein-binarycodesCorrectingDeletions} that VT-codes can correct either a single deletion or insertion. It is also known that the largest VT-codes are obtained for $a=0$, and these codes are conjectured to have the largest cardinality among all single-deletion-correcting codes~\cite{Sloane01onsingle-deletion-correcting}. The redundancy of the $VT_0(n)$ code is at most $\log(n+1)$ (for the  exact cardinality of  the code $VT_0(n)$, see \cite[Eq.~(10)]{Sloane01onsingle-deletion-correcting}). For all $n$, the union of all VT-codes forms a partition of the space $\FTwo^n$, that is $\cup_{a=0}^n VT_a(n) = \FTwo^n$.
\subsubsection{$b$-burst-deletion-correcting codes}
We next review the existing constructions of $b$-burst-deletion-correcting codes, as given in~\cite{ChengSwartFerreiraAbdelGhaffar-ThreeOrMoreBurstDelIns}.

\begin{itemize}
\item Construction~1 from \cite[Section III]{ChengSwartFerreiraAbdelGhaffar-ThreeOrMoreBurstDelIns}: the constructed code is defined to be the set of all codewords $\vec{c}$ such that each row of the $b \times \frac{b}{n}$ array $A_b(\vec{c})$ is a codeword of the code $VT_0(\frac{n}{b})$. A deletion burst of size $b$ deletes exactly one symbol in each row of $A_b(\vec{c})$ which can then be corrected by the VT-code. The redundancy of this construction is $$b\left(\log\left(\frac{n}{b}+1\right)\right) .$$
\item Construction~2 from \cite[Section III]{ChengSwartFerreiraAbdelGhaffar-ThreeOrMoreBurstDelIns}: for every codeword $\vec{c}$ in this construction, the first row of the $b \times \frac{b}{n}$ array $A_b(\vec{c})$ is $(1,0,1,0,\dots)$ (to obtain the position of the deletion of each row to within one symbol). All the other rows are codewords from a code that can correct one deleted bit if it is known to be in one of two adjacent positions.
The redundancy of this construction is $$\frac{n}{b}+(b-1)\log(3).$$
\item Construction~3 from \cite[Section III]{ChengSwartFerreiraAbdelGhaffar-ThreeOrMoreBurstDelIns}: for every codeword $\vec{c}$, the first two rows of the $b \times \frac{b}{n}$ array $A_b(\vec{c})$ are VT-codes together with the property that the run length is at most two. The other rows are again codewords that can correct the deleted bit if it is known to occur in one of two adjacent positions. The redundancy of this construction is approximately:
\begin{align*}
 & 2\frac{n}{b}+(b-2)\log(3) - \log\left(\frac{4\cdot 3^{\frac{n}{b}-1}}{(\frac{n}{b}+1)^2}\right)\\ 
= & \frac{n}{b}+2\log\left(\frac{n}{b}+1\right)+(b-2)\log(3)+ c,
\end{align*}
for some constant $c$.

\end{itemize}

\subsubsection{Correcting a deletion burst of size at most $b$}
To the best of our knowledge, the only known construction to correct a burst of size at most $b$ is the one from~\cite{Bours-InsertionsDeletions}.
Here, encoding is done in an array of size $\frac{n}{b} \times b$ and the stored vector is taken row-wise from the array.
The first $\frac{n}{b}-1$ rows are codewords of a comma-free code (CFC) and the last row is used for the redundancy of an erasure-correcting code (applied column-wise).
Using the size of a CFC from \cite[p.~9]{Bours-InsertionsDeletions}, it is possible to derive that the redundancy of this construction is at least $\frac{n}{b}$ and therefore the code rate is less than one.

\subsubsection{Correcting $b$ deletions (not a burst)}
In \cite{BrakensiekGuruswamiZbarksy-DeletionCorrection}, a construction is presented of codes which correct $b$ deletions at arbitrary positions (not in a burst) in a vector of length $n$. The redundancy of this construction is given by
\begin{equation*}
 c\cdot b^2 \log(b) \log(n),
\end{equation*}
for some constant $c$.

\section{Equivalence of Bursts of Deletions and Bursts of Insertions}\label{sec:equivalence}
In the following, we show the equivalence of bursts of deletions and bursts of insertions. Thus, in the remainder of the paper, whenever we refer to bursts of deletions, all the results hold equivalently for bursts of insertions as well.
\begin{theorem}\label{thm:equiv-burst-del-ins}
A code $\mathcal{C}$ is a $b$-burst-deletion-correcting code if and only if it is a $b$-burst-insertion-correcting code.
\end{theorem}
\begin{IEEEproof}
Note that if $\mathcal{C}$ is a $b$-burst-deletion-correcting code of length $n$, then there are no two vectors in $\FTwo^{n-b}$ which stem from deleting $b$ consecutive symbols in two codewords and are equal.

Now, assume that $\mathcal{C}$ is \emph{not} $b$-burst-insertion-correcting code.
Then, there are two different codewords $\vec{x}$, $\vec{y} \in \mathcal{C}$ of length $n$ such that inserting a $b$-burst in both codewords leads to two equal vectors of length $n+b$. That is, there are two integers $i,j$ (w.l.o.g. $i\leq j$) and two vectors $(s_1,\dots,s_b)$, $(t_1,\dots,t_b)$ such that for $\vec{v} \triangleq(x_1,\dots,x_i,s_1,\dots,s_b,x_{i+1},\dots,x_n)$ and $\vec{w}\triangleq (y_1,\dots,y_j,t_1,\dots,t_b,y_{j+1},\dots,y_n)$, it holds that $\vec{v}=\vec{w}$.

Define a set $\myset{J} =\{i+1,\dots,i+b,j+1,\dots,j+b\}$. If $|\myset{J}| = 2b$, then let $\myset{I} \triangleq \myset{J}$, else $\myset{I} = \myset{J} \cup \{j+b+1,\dots, j+3b-|\myset{J}|\}$ such that in either case $|\myset{I}| = 2b$.

Denote by $\vec{v}_\myset{I}$ and $\vec{w}_\myset{I}$ the two vectors of length $n-b$ that stem from deleting the symbols at the positions in $\myset{I}$ in $\vec{v}$ and $\vec{w}$.
Clearly, $\vec{v}_\myset{I}=\vec{w}_\myset{I}$.
Further, $\vec{v}_\myset{I} =(x_1,\dots,x_{\ell},x_{\ell+b+1},\dots,x_n)$, where $\ell =i $ if $j\leq i+b$ and $\ell = j-b$ else, and $\vec{w}_\myset{I} =(y_1,\dots,y_i,y_{i+b+1},\dots,y_n)$.
However, this is a contradiction since $\vec{x}$ and $\vec{y}$ are codewords of a $b$-burst-deletion-correcting code and thus, the code $\mathcal{C}$ is also a $b$-burst-insertion-correcting code.

The other direction can easily be shown with the same strategy.
\end{IEEEproof}

The proofs of the next two theorems are similar to the one of Theorem~\ref{thm:equiv-burst-del-ins} and thus we omit them.
\begin{theorem}\label{thm:equiv-burst-del-ins-at-most}
A code $\mathcal{C}$ can correct a deletion burst of size at most $b$ if and only if it can correct an insertion burst of size at most $b$.
\end{theorem}
\begin{theorem}\label{thm:equiv-burst-del-ins-at-most-non-cons}
A code $\mathcal{C}$ can correct a non-consecutive deletion burst of size at most $b$ if and only if it can correct a non-consecutive insertion burst of size at most~$b$.
\end{theorem}

\section{An Upper Bound on the Code Size}\label{sec:bound}
The goal of this section is to provide an explicit upper bound on the cardinality of burst-deletion-correcting codes. For large $n$, Levenshtein~\cite{Levenshtein-TwoAdjacentDeletions} derived an asymptotic upper bound on the maximal cardinality of a binary $b$-burst-deletion-correcting code~$\mathcal{C}$ of length $n$. This bound states that for $n$ large enough, an upper bound on the cardinality of the code $\mathcal{C}$ is approximately
\begin{equation*}
\frac{2^{n-b+1}}{n},
\end{equation*}
and hence its redundancy is at least roughly $\log(n) +b -1$.

Our main goal in this section is to provide an explicit upper bound on the cardinality of $b$-burst-deletion-correcting codes. We follow a method which was recently developed by Kulkarni and Kiyavash in~\cite{KulkarniKiyavash-UpperBoundsForDeletion} to obtain such an upper bound.


The size of the $b$-burst-deletion ball for a vector $\vec{x}$ was shown by Levenshtein~\cite{Levenshtein-TwoAdjacentDeletions} to be
\begin{equation}\label{eq:ball_size}
|D_b(\vec{x})| = 1 + \sum\limits_{i=1}^{b} \Big(r( A_b(\vec{x})_i )-1\Big),
\end{equation}
where $r( A_b(\vec{x})_i )$ denotes the number of runs in the $i$-th row of the array $A_b(\vec{x})$. Notice that $1 \leq |D_b(\vec{x})| \leq 1+ (\frac{n}{b}-1) \cdot b = n-b+1$.
\begin{lemma}\label{lem:sizedelball}
Let $\vec{x} \in \FTwo^n$ and $\vec{y}\in \FTwo^{n+b}$ be two vectors such that $\vec{x} \in D_b(\vec{y})$.
Then, $|D_b(\vec{y})| \geq |D_b(\vec{x})|$.  
\end{lemma}  
\begin{IEEEproof}
If $\vec{x} \in D_b(\vec{y})$ then for all $1\leq i\leq b$, $A_b(\vec{x})_i \in D_1(A_b(\vec{y})_i)$, and hence $r(A_b(\vec{x})_i) \leq r(A_b(\vec{y})_i)$, \cite[Lemma~3.2]{KulkarniKiyavash-UpperBoundsForDeletion}. Therefore, according to~(\ref{eq:ball_size}), we get that
\begin{align*}
|D_b(\vec{x})| & = 1 + \sum\limits_{i=1}^{b} \Big(r( A_b(\vec{x})_i )-1\Big) & \\
& \leq 1 + \sum\limits_{i=1}^{b} \Big(r( A_b(\vec{y})_i )-1\Big) = |D_b(\vec{y})|.
\end{align*}


\end{IEEEproof}

We are now ready to provide an explicit upper bound on the cardinality of burst-deletion-correcting codes.
\begin{theorem}
\label{thm:bound}
Any $b$-burst-deletion-correcting code $\mathcal{C}$ of length $n$ satisfies
\begin{equation*}
|\mathcal{C}| \leq \frac{2^{n-b+1}-2^b}{n-2b+1}.
\end{equation*}
\end{theorem}
\begin{IEEEproof}
We proceed similarly to the method presented by Kulkarni and Kiyavash in~\cite[Theorem~3.1]{KulkarniKiyavash-UpperBoundsForDeletion}.
Let $\mathcal{H}_{2,b,n}$ be the following hypergraph:
\begin{equation*}
\mathcal{H}_{2,b,n} = (\FTwo^{n-b}, \{ D_b(\vec{x}) : \vec{x} \in \FTwo^n\} ).
\end{equation*}
The size of the largest $b$-burst-deletion-correcting code equals the matching number of $\mathcal{H}_{2,b,n}$, denoted as in \cite{KulkarniKiyavash-UpperBoundsForDeletion} by $\nu(\mathcal{H}_{2,b,n})$.
By \cite[Lemma~2.4]{KulkarniKiyavash-UpperBoundsForDeletion}, to obtain an upper bound on $\nu(\mathcal{H}_{2,b,n})$, we can construct a fractional transversal, which will give an upper bound on the matching number. The best upper bound according to this method is denoted by $\tau^*(\mathcal{H}_{2,b,n})$ and is calculated according to the following linear programming problem
\begin{align*}
\tau^*(\mathcal{H}_{2,b,n}) &= \min_{w:\FTwo^{n-b}\rightarrow \mathbb{R}} \bigg\{ \sum\limits_{\vec{x} \in \FTwo^{n-b}} w(\vec{x})\bigg\}\\
\text{subject to } \sum\limits_{\vec{x} \in D_b(\vec{y})} w(\vec{x}) &\geq 1, \forall \vec{y} \in \FTwo^n\\
\text{and }\hspace{10ex}w(\vec{x}) &\geq 0, \forall \vec{x} \in \FTwo^{n-b}.
\end{align*}

Next, we will show a weight assignment $w$ to the vectors in $\FTwo^{n-b}$ which provides a fractional transversal. This weight assignment is given by
\begin{equation*}
w(\vec{x}) = \frac{1}{|D_b(\vec{x})|}, \quad \forall \vec{x} \in \FTwo^{n-b},
\end{equation*}
which clearly satisfies that $w(\vec{x}) \geq 0$ for all $\vec{x} \in \FTwo^{n-b}$. Furthermore, according to Lemma~\ref{lem:sizedelball}, we also get that 
for every $\vec{y} \in \FTwo^n$:
\begin{equation*}
\sum_{\vec{x} \in D_b(\vec{y})} w(\vec{x})= 
\sum_{\vec{x} \in D_b(\vec{y})}\frac{1}{|D_b(\vec{x})|} \geq 
\sum_{\vec{x} \in D_b(\vec{y})}\frac{1}{|D_b(\vec{y})|} \geq 
1,
\end{equation*}
and hence $w$ indeed provides a fractional transversal. 

For $1\leq i\leq n-b+1$, let us denote by $N(n,b,i)$ the size of the set $\{ \vec{x}\in \FTwo^{n} : |D_b(\vec{x})| =i \}$. We show in Appendix~\ref{app:N(n,bmi)} that $N(n,b,i) = 2^b\binom{n-b}{i-1}$. The weight of this fractional transversal is given by
\begin{align*}
\sum_{\vec{x}\in \FTwo^{n-b}} w(\vec{x}) &= \sum_{\vec{x}\in \FTwo^{n-b}} \frac{1}{|D_b(\vec{x})|}\\
& = \sum_{i=1}^{n-2b+1} \frac{N(n-b,b,i)}{i}\\
&= 2^b\sum_{i=1}^{n-2b+1} \frac{ \binom{n-2b}{i-1}}{i}\\
&=2^b\sum_{i=1}^{n-2b+1}\frac{ (n-2b)!}{(i-1)!(n-2b-i+1)!i}\\
&=2^b\sum_{i=1}^{n-2b+1}\frac{ (n-2b+1)!}{i!(n-2b-i+1)!(n-2b+1)}\\
&=\frac{2^b}{n-2b+1} \sum_{i=1}^{n-2b+1}\binom{n-2b+1}{i}\\
&=\frac{2^{n-b+1}-2^b}{n-2b+1}.
\end{align*}
Therefore, the value $\frac{2^{n-b+1}-2^b}{n-2b+1}$ is an upper bound on the maximum cardinality of any binary $b$-burst-deletion-correcting code.
\end{IEEEproof}

Notice that for $b=1$ our upper bound in Theorem~\ref{thm:bound} coincides with the upper bound in \cite[Theorem~3.1]{KulkarniKiyavash-UpperBoundsForDeletion} for single-deletion-correcting codes. Furthermore, for $n$ large enough our upper bound matches the asymptotic upper bound from \cite{Levenshtein-TwoAdjacentDeletions}.
Lastly, we conclude that the redundancy of a $b$-burst-deletion-correcting code is lower bounded by the following value
\begin{equation}\label{eq:red_lower_bound}
\log(n-2b+1) - \log(2^{-b+1}-2^{b-n}) \approx \log(n) +b-1.
\end{equation}

\section{Construction of $b$-Burst-Deletion-Correcting Codes}\label{sec:exactly_b}
The main goal of this section is to provide a construction of $b$-burst-deletion-correcting codes, whose redundancy is better than the state of the art results we reviewed in Section~\ref{subsec:rev_work} and is close to the lower bound on the redundancy, which is stated in~(\ref{eq:red_lower_bound}).  We will first explain the main ideas of the construction and will then provide the specific details of the construction.

\subsection{Background}
As shown in Section~\ref{sec:preliminaries}, we will treat the codewords in the $b$-burst-deletion-correcting code as $b\times \frac{n}{b}$ codeword arrays, where $n$ is the codeword length and $b$ divides $n$. Thus, for a codeword $\vec{x}$, the codeword array $A_b(\vec{x})$ is formed by $b$ rows and $\frac{n}{b}$ columns, and the codeword is transmitted column-by-column. Thus, a deletion burst of size $b$ in $\vec{x}$ deletes exactly one bit from each row of the array $A_b(\vec{x})$. That is, if a codeword $\vec{x}$ is transmitted, then the $b\times (\frac{n}{b}-1)$ array representation of the received vector $\vec{y}$ has the following structure 
	\[ A_b(\vec{y})=\begin{bmatrix}
	y_1 & y_{b+1} &  \dots  & y_{n-2b+1} \\
	y_2 & y_{b+2} & \dots  & y_{n-2b+2} \\
	\vdots & \vdots & \ddots & \vdots \\
	y_b & y_{2b} & \dots  & y_{n-b}
	\end{bmatrix}. \]
Each row is received by a single deletion of the corresponding row in $A_b(\vec{x})$~\cite{ChengSwartFerreiraAbdelGhaffar-ThreeOrMoreBurstDelIns}, i.e., $A_b(\vec{y})_i\in D_1(A_b(\vec{x})_i)$, $\forall 1\leq i\leq b$.

Since the channel deletes a burst of \textit{b} bits, the deletions can span at most two columns of the codeword array. Therefore, information about the position of a deletion in a single row provides information about the positions of the deletions in the remaining rows. However, note that deletion-correcting codes are not always able to determine the exact position of the deleted bit. For example, assume the all-zero codeword was transmitted and a single deletion of one of the bits has occurred. Even if the decoder can successfully decode the received vector, it is not possible to know the position of the deleted bit since it could be any of the bits. 

In order to take advantage of the correlation between the positions of the deleted bits in different rows and overcome the difficulty that deletion-correcting codes cannot always provide the location of the deleted bits, we construct a single-deletion-correcting code with the following special property. The receiver of this code can correct the single deletion and determine its location within a certain predetermined range of consecutive positions. This code will be used to encode the first row of the codeword array and will provide partial information on the position of the deletions for the remaining $b-1$ rows. In these rows, we use a different code that will take advantage of this positional information.

The following is a high-level outline of the proposed codeword array construction:
\begin{itemize}
\item The first row in the array is encoded as a VT-code in which we restrict the longest run of 0's or 1's to be at most $ \log(2n) $. The details of this code are described in Section~\ref{subsec:rll}.
\item Each of the remaining $ (b-1) $ rows in the array is encoded using a modified version of the VT-code, which will be called a \textit{shifted VT} (\textit{SVT})-\textit{code}. This code is able to correct a single deletion in each row once the position where the deletion occurred is known to within $ \log(2n)+1 $ consecutive positions. The details of these codes are discussed in Section~\ref{subsec:svt}.
\end{itemize}
Section~\ref{subsec:full} presents the full code construction. Let us explore the different facets of our proposed codeword array construction in more detail.

\subsection{Run-length Limited (RLL) VT-Codes}\label{subsec:rll}
In general, a decoder for a VT-code can decode a single deletion while determining only the position of the run that contains the deletion, but not the exact position of the deletion itself. For this reason, we seek to limit the length of the longest run in the first row of the codewords array.

A length-$n$ binary vector is said to satisfy the \emph{$(d,k)$ Run Length Limited (RLL)} constraint, denoted by $RLL_n(d,k)$, if between any two consecutive 1's there are at least $d$ 0's and at most $k$ 0's~\cite{Immink91}. Since we are concerned with runs of 0's or 1's, we will state our constraints on the longest runs of 0's and 1's. Note that the maximum rate of codes which satisfy the $(d,k)$ RLL constraint for fixed $d$ and~$k$ is less than 1. To achieve codes with asymptotic rate 1, the restriction on the longest run is a function of the length $n$. 
\begin{definition}
	A length-$n$ binary vector $\vec{x}$ is said to satisfy the $\textbf{f(n)}$\textbf{-RLL(n)} constraint, and is called an $\textbf{f(n)}$\textbf{-RLL(n)} vector, if the length of each run of 0's or 1's in $\vec{x}$ is at most $f(n)$. 
\end{definition}
A set of $f(n)$-RLL$(n)$ vectors is called an \emph{$f(n)$-RLL$(n)$ code}, 
and the set of all $f(n)$-RLL$(n)$ vectors is denoted by $S_n(f(n))$. The \emph{capacity} of the $f(n)$-RLL$(n)$ constraint is  
$$C(f(n)) = \limsup_{n\rightarrow \infty}\frac{\log (|S_n(f(n))|)}{n},$$
and for the case in which the capacity is 1, we define also the \emph{redundancy} of the $f(n)$-RLL$(n)$ constraint to be 
$$r(f(n)) = n- \log (|S_n(f(n))|).$$
\begin{lemma}\label{lem:RLL_red}
The redundancy of the $ \log(2n) $-RLL(n) constraint is upper bounded by 1 for all $n$, and it asymptotically approaches $ \log(e)/2 \approx 0.36 $. 
\end{lemma}
\begin{IEEEproof}
For simplicity let us assume that $n$ is a power of two. Let $X_n$ be a random variable that denotes the length of the longest run in a length-$n$ binary vector, where the vectors are chosen uniformly at random. We will be interested in computing a lower bound on the probability $$P(X_n\leq \log(2n))=P(X_n\leq 1+\log(n)),$$ or an upper bound on the probability $P(X_n \geq 2+\log(n))$. 
By the union bound it is enough to require that every window of $2+\log(n)$ bits is not all zeros or all ones and thus we get that
$$P(X_n \geq  2+\log(n)) \leq n\cdot \frac{2}{2^{2+\log(n)}} = \frac{1}{2},$$
and thus $P(X_n\leq 1+ \log(n)) \geq 1/2$. Therefore the size of the set $S_n(\log(2n))$ is at least $2^n/2$ and its redundancy $r(\log(2n))$ is at most one bit. 

In order to find the asymptotic behavior of $r(\log(2n))$, we use the following result from~\cite{schilling1990longest}. Let $Y_n$ be a random variable that denotes the length of the longest run of ones in a length-$n$ binary vector which is chosen uniformly at random, and $W$ is a continuous random variable whose cumulative distribution function is given by $F_W(x) = e^{-(1/2)^{x}}$. Then, the following holds:
\begin{align*}
& P(X_n\leq \log(n)+1) = P(Y_{n-1}\leq \log (n)) & \\
\approx & P\left(W\leq \log(n)+1-\log\left(\frac{n-1}{2}\right)\right)& \\
= & P\left(W\leq \log\left(\frac{n}{n-1}\right)+2\right) & \\
= & e^{-(1/2)^{\log\left(\frac{n}{n-1}\right)+2}} = e^{-(1/4)\cdot \frac{n-1}{n}} = \left(\frac{1}{{e^{1/4}}}\right)^{1-\frac{1}{n}}.&
\end{align*}
Therefore, for $n$ large enough $P(X_n\leq \log(n)+1)\approx e^{-1/4}$, and $r(\log(2n))\approx \log(e)/4 \approx 0.36$.
\end{IEEEproof}
\begin{remark}
Since $ \log(e)/2 < 1 $, we can guarantee that the encoded vector will not have a run of length longer than $ \log(2n) $ with the use of a single additional redundancy bit. Thus $ \log(2n) $ is a proper choice for our value of $f(n)$; a smaller $ f(n) $ would substantially increase the redundancy of the first row, and a larger $ f(n) $ would not help since setting $ f(n)=\log(2n) $ already only requires at most a single bit of redundancy. Note that Lemma \ref{lem:RLL_red} agrees with the results from~\cite{renyi1970probability,schilling1990longest} which state that the typical length of the longest run in \textit{n} flips of a fair coin converges to $ \log(n) $. Lastly we note that in Appendix~\ref{RLL_encoding}, we provide an algorithm to efficiently encode/decode run-length-limited sequences for the $(\log(n)+3)$-RLL$(n)$ constraint.
\end{remark}

Recall that our goal was to have the vector stored in the first row be a codeword in a VT-code so it can correct a single deletion and also limit its longest run. Hence we define a family of codes which satisfy these two requirements by considering the intersection of a VT-code with the set $S_n(f(n))$.
\begin{definition}
Let $a,n$ be two positive integers where $0\leq a\leq n$. The $VT_{a,f(n)}(n)$ code is defined to be the intersection of the codes $VT_{a}(n)$ and $S_n(f(n))$. That is,
$$ VT_{a,f(n)}(n) = \bigg\{\vec{x}\ : \ \vec{x} \in VT_{a}(n), \vec{x} \in S_n(f(n))  \bigg\}.$$
\end{definition}

Note that since $ VT_{a,f(n)}(n)$ is a subcode of $VT_{a}(n)$, it is also a single-deletion-correcting code. The following lemma is an immediate result on the cardinality of these codes.
\begin{lemma}\label{RLL_VT_red}
For all $n$, there exists $0\leq a\leq n$ such that 
$$|VT_{a,f(n)}(n)|\geq \frac{|S_n(f(n))|}{n+1}.$$
\end{lemma}
\begin{IEEEproof}
The VT-codes form a partition of $\F_2^n$ into $n+1$ different codebooks $VT_0(n),VT_1(n),\ldots, VT_n(n)$. Using the pigeonhole principle, we can determine the lower bound of the maximum intersection between these $n+1$ codebooks and $S_n(f(n))$ and get that
\[  \max_{0\leq a \leq n}\bigg\{|S_n(f(n)) \cap VT_a(n)| \bigg\}\geq \dfrac{|S_n(f(n))|}{n+1}. \]
\end{IEEEproof}
We conclude with the following corollary.
\begin{corollary}\label{cor:RLL_VT_red}
For all $n$, there exists $0\leq a\leq n$ such that the redundancy of the code $VT_{a,\log(2n)}(n)$ is at most $\log(n+1) + 1$ bits.
\end{corollary}

\subsection{Shifted VT-Codes}\label{subsec:svt}
Let us now focus on the remaining $ (b-1) $ rows of our codeword array. Decoding the first row in the received array allows the decoder to determine the locations of the deletions of the remaining rows up to a set of consecutive positions. We define a new class of codes with this positional knowledge of deletions in mind.

\begin{definition}\label{def:bounded deletion codes}
A \textbf{P-bounded single-deletion-correcting code} is a code in which the decoder can correct a single deletion given knowledge of the location of the deleted bit to within $ P $ consecutive positions.
\end{definition}

We create a new code, called a \emph{shifted VT} (\emph{SVT})-\emph{code}, which is a variant of the VT-code and is able to take advantage of the positional information as defined in Definition~\ref{def:bounded deletion codes}.
\begin{construction}\label{const:svt}
	For $0\leq c < P$ and $d\in\{0,1\}$, let the shifted Varshamov-Tenengolts code $ SVT_{c,d}(n,P) $ be:
	\[ SVT_{c,d}(n,P) \hspace{-0.5ex} \triangleq \hspace{-0.5ex} \bigg\{\!  \vec{x} : \sum_{i=1}^n\hspace{-0.25ex} i x_i \equiv c~(\bmod P) , \sum_{i=1}^n \hspace{-0.25ex} x_i \equiv d~(\bmod 2)   \bigg\}  . \]
\end{construction}
Other modifications of the VT-code have previously been proposed in~\cite{cullina2012coloring} to improve the upper bounds on the cardinality of deletion-correcting codes. The next lemma proves the correctness of this construction and provides a lower bound on the cardinality of these codes.
\begin{lemma}\label{lem:svt_p}
For all $0\leq c <P$ and $d\in\{0,1\}$, the $SVT_{c,d}(n,P) $-code (as defined in Construction~\ref{const:svt}) is a P-bounded single-deletion-correcting code.
\end{lemma}
\begin{IEEEproof}
In order to prove that the $SVT_{c,d}(n,P) $-code is a $P$-bounded single-deletion-correcting code, it is sufficient to show that there are no two codewords $\vec{x},\vec{y} \in SVT_{c,d}(n,P)$ that have a common subvector of length $n-1$ where the locations of the deletions are within $P$ positions.

Assume in the contrary that there exist two different codewords $\vec{x},\vec{y} \in SVT_{c,d}(n,P)$, where there exist $1\leq k,\ell \leq n$, where $|\ell-k| < P$, such that $\vec{z} = \vec{x}_{[n]\setminus \{k\}} = \vec{y}_{[n]\setminus \{\ell\}}$, and assume that $k<\ell$. Since $\vec{x},\vec{y} \in SVT_{c,d}(n,P)$, we can summarize these assumptions in the following three properties:
\begin{enumerate}
	\item $ \sum_{i=1}^{n}x_i - \sum_{i=1}^{n}y_i \equiv 0~(\bmod 2) $.
	\item $ \sum_{i=1}^{n}ix_i - \sum_{i=1}^{n}iy_i \equiv 0~(\bmod P) $.
	\item $ \ell-k<P $.
\end{enumerate}

According to these assumptions and since $\vec{x}_{[n]\setminus \{k\}} = \vec{y}_{[n]\setminus \{\ell\}}$, it is evident that $k$ is the smallest index for which $x_k \neq y_k $, and $\ell$ is the largest index for which $x_\ell \neq y_\ell$. Additionally, from the first property $\vec{x}$ and $\vec{y}$ have the same parity and thus $x_k = y_\ell$. Outside of the indices $k$ and $\ell$, $\vec{x}$ and $\vec{y}$ are identical, while inside they are shifted by one position: 
\begin{align*}
	& x_i = y_i  \text{ \hspace{2ex}  for } i<k \text{ and } i>\ell, & \\
	& x_i = y_{i-1}  \text{\hspace{0.5ex}  for } k<i\leq \ell. &
\end{align*}
We consider two scenarios: $x_k=y_\ell=0$ or $x_k=y_\ell=1 $.
First assume that $x_k=y_\ell=0$, and in this case we get that
\begin{align*}
	&\sum\limits_{i=1}^{n}ix_i - \sum\limits_{i=1}^{n}iy_i = \sum\limits_{i=k}^{\ell}ix_i - \sum\limits_{i=k}^{\ell}iy_i = \sum\limits_{i=k+1}^{\ell}ix_i - \sum\limits_{i=k}^{\ell-1}iy_i &\\
	&= \sum\limits_{i=k+1}^{\ell}iy_{i-1} - \sum\limits_{i=k}^{\ell-1}iy_i = \sum\limits_{i=k}^{\ell-1}(i+1)y_{i} - \sum\limits_{i=k}^{\ell-1}iy_i  = \sum\limits_{i=k}^{\ell-1}y_i. &
\end{align*}
The sum  $\sum_{i=k}^{\ell-1}y_i$ cannot be equal to zero or else we will get that $\vec{x}=\vec{y}$, and hence 
$$0 < \sum\limits_{i=1}^{n}ix_i - \sum\limits_{i=1}^{n}iy_i = \sum\limits_{i=k}^{\ell-1}y_i \leq \ell - k < P,$$
in contradiction to the second property.

A similar contradiction can be shown for $x_k=y_\ell=1$. Thus, the three properties cannot all be true, and the $SVT_{c,d}(n,P) $-code is a $P$-bounded single-deletion-correcting code.
\end{IEEEproof}
\begin{lemma}\label{lem:red_svt}
There exist $0\leq c < P$ and $d\in\{0,1\}$ such that the redundancy of the $SVT_{c,d}(n,P)$ code as defined in Construction~\ref{const:svt} is at most $\log(P)+1$ bits.
\end{lemma}
\begin{IEEEproof}
Similarly to the partitioning of the VT-codes, the $2P$ codes $SVT_{c,d}(n,P)$, for $0\leq c < P$ and $d\in\{0,1\}$, form a partition of all length-$n$ binary vectors into $2P$ mutually disjoint sets. Using the pigeonhole principle, there exists a code whose cardinality is at least $ \frac{2^n}{2P} $ and thus its redundancy is at most $ \log(2P) = \log(P)+1 $ bits.
\end{IEEEproof}

There are two major differences between the SVT-codes and the usual VT-codes. First, the SVT-codes restrict the overall parity of the codewords. This parity constraint costs an additional redundancy bit, but it allows us to determine whether the deleted bit was a 0 or a 1. Second, in the VT-code, the weights assigned to each element in the vector are $1,2,\ldots,n $; on the other hand, in the SVT-code, these weights can be interpreted as repeatedly cycling through $1,2,\ldots,P-1,0$ (due to the $(\bmod P)$ operation). Because of these differences, a VT-code requires roughly $\log(n+1) $ redundancy bits while a SVT-code requires approximately only $\log(P)+1$ redundancy bits.

The proof of Lemma~\ref{lem:svt_p} motivates also the operation of a decoder to the SVT-code. In order to complete the description of this code we show in Appendix~\ref{SVT_decoder} the full description of this decoder for the SVT-codes.

\subsection{Code Construction}\label{subsec:full}
We are now ready to construct $b$-burst-deletion-correcting codes by combining the ideas from the previous two subsections into a single code. 

\begin{construction}\label{const:cons_burst}
Let $\mathcal{C}_1$ be a $VT_{a,\log(2n/b)}(n/b)$ code for some $0\leq a\leq n/b$ and let $\mathcal{C}_2$ be a shifted VT-code $SVT_{c,d}(n/b,\log(n/b)+2)$ for $0\leq c < n/b+2$ and $d\in\{0,1\}$. 
The code $\mathcal{C}$ is constructed as follows
$$\mathcal{C} \triangleq \{ \vec{x} :  A_b(\vec{x})_1\in \mathcal{C}_1, A_b(\vec{x})_i\in \mathcal{C}_2, \textrm{ for $2\leq i\leq b$}  \}.$$
\end{construction}

\begin{theorem}
	The code $\mathcal{C}$ from Construction~\ref{const:cons_burst} is a $b$-burst-deletion-correcting code.
\end{theorem}
\begin{IEEEproof}
	Assume $\vec{x}\in\mathcal{C}$ is the transmitted vector and $\vec{y}\in D_b(\vec{x})$ is the received vector. In the $b\times (n/b-1)$ array $A_b(\vec{y})$, every row is therefore received by a single deletion of the corresponding row in $A_b(\vec{x})$. 
	
	Since the first row of $A_b(\vec{x})_1$ belongs to a $VT_{a,\log(2n/b)}(n/b)$ code, the decoder of this code can successfully decode and insert the deleted bit in the first row of $A_b(\vec{y})_1$. Furthermore, since every run in $A_b(\vec{x})_1$ consists of at most $\log(2n/b)$ bits, the locations of the deleted bits in the remaining rows are known within $\log(n/b)+2$ consecutive positions. 
	Finally, the remaining $b-1$ rows decode their deleted bit since they belong to a shifted VT-code $SVT_{c,d}(n/b,\log(n/b)+2)$ (Lemma~\ref{lem:svt_p}).
\end{IEEEproof}

To conclude this discussion, the following corollary summarizes the result presented in this section.
\begin{corollary}
	For sufficiently large $ n $, there exists a $b$-burst-deletion-correcting code whose number of redundancy bits is at most
	\begin{align*}
	\log(n) + (b-1)\log(\log(n)) +b -\log(b).
	\end{align*}
\end{corollary}

\section{Correcting a Burst of Length at most $b$ (consecutively)}\label{sec:at_most_b}
In this section, we consider the problem of correcting a burst of consecutive deletions of length at most $b$. As defined in Section~\ref{sec:preliminaries}, a code capable of correcting a burst of at most $ b $ consecutive deletions needs to be able to correct any burst of size $a$ for $a\leq b$. For the remainder of this section, we assume that $(b!) | n$.

The case $b=2$ was already solved by Levenshtein with a construction that corrects a single deletion or a deletion of two adjacent bits~\cite{Levenshtein-TwoAdjacentDeletions}. The redundancy of this code, denoted by $ \mathcal{C}_L(n) $, is at most $1+\log(n)$ bits. Hence this code asymptotically achieves the upper bound for correcting a burst of exactly $2$ deletions.

The general strategy we use in correcting a burst of length \textit{at most} $ b $ is to construct a code from the intersection of the code $\mathcal{C}_L(n)$ with the codes that correct a burst of length \textit{exactly} $ i $, for $3 \leq i \leq b$. 
We refer to each $ i $ as a \textit{level} and in each level we will have a set of codes which forms a partition of the space. Thus, our overall code will be the largest intersection of the codes at each level.

Let us first introduce a simple code construction that can be used as a baseline comparison. We use Construction 1 from~\cite{ChengSwartFerreiraAbdelGhaffar-ThreeOrMoreBurstDelIns}, which is reviewed in Section~\ref{subsec:rev_work}, to form the code in each level $3\leq i\leq b$. Note that in each level we can have a family of codes which forms a partition of the space. Then, the intersection of the codes in each level together with $\mathcal{C}_L(n) $ forms a code that corrects burst of consecutive deletions of length at most $ b $. 

As we mentioned above, the redundancy of the code $\cC_L(n)$ is $\log (n)+1$ and it partitions the space into $ 2n $ codebooks. Similarly, for $3\leq i\leq b$, the redundancy of the codes from~\cite{ChengSwartFerreiraAbdelGhaffar-ThreeOrMoreBurstDelIns} in the $i$th level is $i\left(\log(n/i+1) \right)$, and they partition the space into
$\left( \frac{n}{i}+1 \right)^i $ codebooks. Therefore, we can only claim that the redundancy of this code construction will be approximately 
$$\log (2n) + \sum_{i=3}^b \hspace{-0.3ex}i\left(\log\hspace{-0.3ex}\left(\frac{n}{i}\hspace{-0.3ex}+\hspace{-0.3ex}1\right)\hspace{-0.3ex} \right)\hspace{-0.5ex}\geq \hspace{-0.3ex}\left(\hspace{-0.3ex}\binom{b}{2}\hspace{-0.3ex}-2\hspace{-0.3ex}\right) \hspace{-0.3ex} \log(n) -  \log\left(\prod_{i=2}^bi!\right).$$
Let us denote this simple construction, which provides a baseline redundancy, as $ \mathcal{C}_B(n) $.

The approach we take in this section is to build upon the codes we develop in Section~\ref{sec:exactly_b} and leverage them as the codes in each level instead of the ones from~\cite{ChengSwartFerreiraAbdelGhaffar-ThreeOrMoreBurstDelIns}. However, since the codes from Section~\ref{sec:exactly_b} do not provide a partition of the space we will have to make one additional modification in their construction so it will be possible to intersect the codes in each level and get a code which corrects a burst of size at most $b$.

Recall that in our code from Construction~\ref{const:cons_burst} we needed the first row in our codeword array, $A_b(\vec{x})_1$, to be run-length limited so that the remaining rows could effectively use the SVT-code. Similarly, in order to correct at most \textit{b} consecutive deletions we want the first row of each level's codeword array to be an $N_b$-RLL$(\frac{n}{i})$-vector, where $N_b = \lceil\log(n\log(b))\rceil+1$. In other words, $A_i(\vec{x})_1$ will satisfy the $N_b$-RLL$(\frac{n}{i})$ constraint for $ 3 \leq i \leq b $. Note that the $f(n)$-RLL$(\frac{n}{i})$ constraint does not depend on $i$. We add the term \textit{universal} to signify that an RLL constraint on a vector refers to the RLL constraint on the first row of each level.

\begin{definition}
	A length-$n$ binary vector $\vec{x}$ is said to satisfy the $\textbf{f(n)}$\textbf{-URLL$(n,b)$} constraint, and is called an $\textbf{f(n)}$\textbf{-URLL$(n,b)$} vector, if the length of each run of 0's or 1's in $A_i(\vec{x})_1$ for $ 3\leq i \leq b $, is not greater than $f(n)$. Additionally, the set of all $\textbf{f(n)}$\textbf{-URLL$(n,b)$} vectors is denoted by $ U_{n,b}(f(n)) $.
\end{definition}

We define the \emph{redundancy} of the $f(n)$-URLL$(n,b)$ constraint to be \[r_U(f(n)) = n- \log (|U_{n,b}(f(n))|). \]

\begin{lemma}\label{lem:URLL_red}
The redundancy of the $ N_b $-URLL(n,b) constraint is upper bounded by $ \log(\log(b))-1$ bits: 
\[r_U(N_b) \leq \log(\log(b))-1. \]
\end{lemma}
\begin{IEEEproof}
Using the union bound, we can derive an upper bound on the percentage of sequences in which $A_i(\vec{x})_1$ does not satisfy the $ N_b $-RLL$ (\frac{n}{i}) $ constraint for $3\leq  i \leq b $.
\begin{align*}
\dfrac{|\{\vec{x} :  A_i(\vec{x})_1 \notin S_{\frac{n}{i}}(N_b)\}|}{2^n} &\leq\frac{n}{i}\cdot \left(\frac{1}{2}\right)^{N_b-1}\\
&= \frac{n}{i}\cdot \left(\frac{1}{2}\right)^{\lceil\log(n\log(b))\rceil}\\
&\leq \frac{n}{in\log(b)}\\
&= \frac{1}{i\log(b)}.
\end{align*}
Using the previous result we find an upper bound on the percentages of sequences which do not satisfy the universal RLL constraint.
\begin{align*}
\dfrac{|\{\vec{x} :  \vec{x} \notin U_{n,b}(N_b)\}|}{2^n} &\leq \sum_{i=3}^{b}\left( \frac{1}{i\log(b)} \right)\\ 
&=\left( \frac{1}{\log(b)} \right)\sum_{i=3}^{b}\left( \frac{1}{i}\right)\\
&< \left(\frac{1}{\log(b)}\right)(\ln(b)-2)\\
&= 1-\frac{2}{\log(b)},
\end{align*}
where the last inequality holds since $\sum_{i=1}^n (1/i) < \ln(n) + 1$, for all $n$.
Therefore, we can lower bound the total number of sequences that meet our universal RLL-constraint by:
\begin{align*}
|\{\vec{x} : \vec{x} \in U_{n,b}(N_b)\}| &> 2^n\left[ 1- \left( 1-\frac{2}{\log(b)} \right) \right] \\
&=\frac{2^{n+1}}{\log(b)}.
\end{align*}
Finally, we derive an upper bound on the redundancy of the set $U_{n,b}(N_b)$ to be
\begin{align*}
r_U(N_b) &= n-\log(|U_{n,b}(N_b)|)\\
& < n - \log\left( \frac{2^{n+1}}{\log(b)} \right)\\
&= n - (n+1) + \log(\log(b))\\
&= \log(\log(b)) -1.
\end{align*}
\end{IEEEproof}

In addition to limiting the longest run in the first row of every level, each vector $A_i(\vec{x})_1$ should be able to correct a single deletion. We define the following family of codes.
\begin{construction}\label{con:at most consecutive}
	Let $n$ be a positive integer and $\vec{a}=a_3,\ldots, a_b$ a vector of non-negative integers such that $0\leq a_i\leq n/i$ for $3\leq i\leq b$. The code $\bar{VT}_{\vec{a},f(n)}(n)$ code is defined as follows:
	\begin{equation*}
	\begin{split}
	\bar{VT}_{\vec{a},f(n)}(n) \triangleq \bigg\{\vec{x}\ : \ & A_i(\vec{x})_1 \in VT_{a_i}\left(\frac{n}{i}\right), 3\leq i \leq b,\\
	&\vec{x} \in U_{n,b}(f(n))  \bigg\}.
	\end{split}
	\end{equation*}
\end{construction}

\begin{lemma}\label{URLL_VT_red}
For all $n$, there exists vector $\vec{a}=(a_3,\ldots, a_b)$ such that $0\leq a_i\leq n/i$ for all $3\leq i\leq b$ and 
\[ |\bar{VT}_{\vec{a},f(n)}(n)|\geq \frac{|U_{n,b}(f(n))|}{n^{b-2}} \]
\end{lemma}
\begin{IEEEproof}
For $3\leq i\leq b$, the VT-code $VT_{a_i}\left(\frac{n}{i}\right)$ for $A_i(\vec{x})_1$ forms a partition of all length-$n$ binary sequences into $\frac{n}{i}+1$ different codebooks. Using the pigeonhole principle, we can determine the lower bound of the maximum intersection between the $\frac{n}{i}+1$ codebooks on each level and $U_n(f(n))$ to get
	\begin{align*}
	\max_{\vec{a}}\bigg\{|\bar{VT}_{\vec{a},f(n)}(n)| \bigg\}&=\dfrac{|U_{n,b}(f(n))|}{\prod_{i=3}^{b}\left(\frac{n}{i}+1\right)}\\
	&\geq \dfrac{|U_{n,b}(f(n))|}{n^{b-2}}
	\end{align*}
\end{IEEEproof}

We combine Lemma~\ref{lem:URLL_red} and Lemma~\ref{URLL_VT_red} to find the total redundancy required to satisfy our conditions for the first rows in the codeword arrays. To simplify notation, in the rest of this section whenever we refer to a vector $\bfa$ we refer to $\bfa=(a_3,\ldots,a_b)$ where $0\leq a_i\leq n/i$ for $3\leq i\leq b$.
\begin{corollary}\label{cor:URLL_VT_red}
	For all $n$, there exists a vector $\bfa=(a_3,\ldots,a_b)$ such that the redundancy of the code $\bar{VT}_{\vec{a},N_b}(n)$ is at most $(b-2)\log(n) + \log(\log(b))$ bits.
\end{corollary}

With the universal RLL-constraint in place, we can use the SVT-codes defined in Section~\ref{sec:exactly_b} for each of the remaining rows in each level.

\begin{construction}\label{const:cons_at_most}
Let $\mathcal{C}_L(n)$ be the code from~\cite{Levenshtein-TwoAdjacentDeletions}, $\mathcal{C}_1$ be the code $\bar{VT}_{\vec{a},N_b}(n)$ for some vector $\vec{a}$, and for $3\leq i\leq b$ let $\mathcal{C}_{2,i}$ be a shifted VT-code $SVT_{c_i,d_i}(n/i,N_b+1)$ for $0\leq c_i\leq n/i$ and $d_i\in\{0,1\}$. 
	The code $\mathcal{C}$ is constructed as follows
	\begin{align*}
	\mathcal{C} \triangleq\{\vec{x} : \ &\vec{x} \in \mathcal{C}_L(n), \vec{x} \in \mathcal{C}_1\\
	&A_i(\vec{x})_j \in \mathcal{C}_{2,i}, \textrm{ for } 3\leq i \leq b, 2 \leq j \leq i\}.
	\end{align*}
\end{construction}

\begin{theorem}\label{thm:at_most_correctness}
The code $ \mathcal{C} $ from Construction~\ref{const:cons_at_most} can correct any consecutive deletion burst of size at most $b$. 
\end{theorem}
\begin{IEEEproof}
	Assume $\vec{x}\in\mathcal{C}$ is the transmitted vector and $\vec{y}\in D_i(\vec{x})$ is the received vector, $ 0 \leq i \leq b $. First, by the length of $\vec{y}$ we can easily determine the value of $i$. Recall that the received vector $\vec{y}$ can be represented by an $i\times (n/i-1)$ array $A_i(\vec{y})$ in which every row is received by a single deletion of the corresponding row in $A_i(\vec{x})$. 
	
Since the first row $A_i(\vec{x})_1$ belongs to a $\bar{VT}_{\vec{a},N_b}(n)$ code, the decoder of this code can successfully decode and insert the deleted bit in the first row of $A_i(\vec{y})$. Furthermore, since every run in $A_i(\vec{x})_1$ consists of at most $N_b$ bits, the locations of the deleted bits in the remaining rows are known within $N_b+1$ consecutive positions. 
	Finally, the remaining $i-1$ rows decode their deleted bit since they belong to a shifted VT-code $SVT_{c_i,d_i}(n/i,N_b+1)$ (Lemma~\ref{lem:svt_p}).
\end{IEEEproof}

To conclude, we calculate the amount of redundancy bits needed for Construction~\ref{const:cons_at_most}.

\begin{corollary}\label{cor:at_most_red}
For sufficiently large $ n $, there exists a code which can correct a consecutive deletion burst of size at most $b$ whose number of redundancy bits is at most
\begin{align*}
(b-1)\log(n) \!+\!\left( \binom{b}{2}-1 \right)\log(\log(n)) +\!  \binom{b}{2}+ \log(\log(b)).&
\end{align*}
\end{corollary}
\begin{IEEEproof}
As previously noted, the code $ \mathcal{C}_L(n) $ requires $ \log(n)+1 $ redundancy bits. Corollary~\ref{cor:URLL_VT_red} yields the total number of redundancy bits required for $ \mathcal{C}_1 $. For each level $ i $, $ 3 \leq i \leq b $, there are $ i-1 $ rows we encode with an SVT-code, which yields $ \binom{b}{2}-1 $ total rows. The redundancy for the SVT-code is given by Lemma~\ref{lem:red_svt}.
\end{IEEEproof}

Note that Corollary~\ref{cor:at_most_red} yields a redundancy substantially lower than the redundancy required for the baseline comparison code $\mathcal{C}_B(n)$. In the latter code the $ \log(n) $ redundancy term is quadratic in $ b $, while in the redundancy in Corollary~\ref{cor:at_most_red} the $ \log(n) $ term is linear in $ b $.

\section{Correcting a Burst of Length at most $b$ (non-consecutively)}\label{sec:non_consecutive}
In this section, we will describe a construction for correcting a non-consecutive deletion burst of length at most $b$ for $b \leq 4$. Note that for $b=1$, we can use a VT-code and for $b=2$, we use Levenshtein's construction~\cite{Levenshtein-TwoAdjacentDeletions}. The construction uses a code which can correct two deletions immediately followed by an insertion. For the remainder of this section, we assume that $(b!) | n$.

\subsection{A 2-Deletion-1-Insertion-Burst Correcting Code}\label{subsec:2D-1I}
This subsection describes a code that corrects a deletion burst of size $2$ followed by an insertion at the same position. For shorthand, we refer to this type of error as a \emph{$(2,1)$-burst}, such a code is called a \emph{$(2,1)$-burst-correcting code}, and the set of all $(2,1)$-bursts of a vector $\vec{x}$ is denoted by $D_{2,1}(\vec{x})$. For instance, if the vector $\vec{x}=(0,1,0,0,1,0) \in \mathbb{F}_2^6$ is transmitted then the set of possible received sequences given that a single $(2,1)$-burst occurs to $\vec{x}$ is 
\begin{align*}
D_{2,1}(\vec{x}):=\{ &( \textcolor{red}{0},0,0,1,0), (\textcolor{red}{1},0,0,1,0), (0,\textcolor{red}{1},0,1,0),\\ &(0,1,\textcolor{red}{1},1,0), 
(0,1,0,\textcolor{red}{0},0), (0,1,0,0,\textcolor{red}{1}) \}.
\end{align*}
Note that $D_{1}(\vec{x}) \subseteq D_{2,1}(\vec{x})$ and hence every $(2,1)$-burst-correcting code is a single-deletion-correcting code as well.

We now introduce a construction for $(2,1)$-burst-correcting codes.
\begin{construction}\label{constr:2-1-Code}
For three integers $n \geq 4$, $a \in \mathbb{Z}_{2n-1}$, and $c \in \mathbb{Z}_4$, the code $\cC_{2,1}(n,a,c)$ is defined as follows:
\begin{align*}
\mathcal{C}_{2,1}(n,a,c) \triangleq\   \Big \{ \vec{x} \in  \F_2^n : & \sum_{i=1}^n x_i \equiv c~(\bmod 4), \\
& \sum_{i=1}^n i \cdot x_i \equiv a~(\bmod (2n-1)) \Big \}.
\end{align*}
\end{construction}
Notice that $\cC_{2,1}(n,a,c)$ is a single-deletion-correcting code \cite{Levenshtein-binarycodesCorrectingDeletions}.

In order to prove the correctness of this construction, we introduce some additional terminology. For $(b_1, b_2) \in \mathbb{F}_2^2$, $a \in \mathbb{F}_2$, and $\vec{x} \in \mathbb{F}_2^n$ let $D_{2,1}(\vec{x})^{(b_1,b_2) \to a} \subseteq D_{2,1}(\vec{x})$ be the set of vectors from $D_{2,1}(\vec{x})$ that result from the deletion of the subvector $(b_1, b_2)$ followed by the insertion of $a$. For example, for the vector $\vec{x}=(0,1,0,0,0,1,0)$, 
\begin{align*}
&D_{2,1}^{(0,0) \to 1} (\vec{x})=\{ (0,1,\textcolor{red}{1},0,1,0),(0,1,0,\textcolor{red}{1},1,0) \},&\\
&D_{2,1}^{(0,0) \to 0} (\vec{x})=\{ (0,1,0,0,1,0) \}.&
\end{align*}

The following claim follows in a straightforward manner.

\begin{claim}\label{claim:2-1burst}
For any $(a,b_1,b_2) \not \in \{ (1,0,0), (0,1,1) \}$ $D_{2,1}^{(b_1, b_2) \to a}(\vec{x}) \subseteq D_1(\vec{x})$.\end{claim}

We are now ready to prove the correctness of Construction~\ref{constr:2-1-Code}.
\begin{theorem} 
Let $n \geq 4$, $a \in \mathbb{Z}_{2n-1}$, and $c \in \mathbb{Z}_4$ be three integers. Then, the code $\cC_{2,1}(n,a,c)$ from Construction~\ref{constr:2-1-Code} is a $(2,1)$-burst-deletion correcting code.
\end{theorem}
\begin{IEEEproof}
We will show that for all $\vec{x}, \vec{y} \in \cC_{2,1}(n,a,c)$,  $\cD_{2,1}(\vec{x}) \cap \cD_{2,1}(\vec{y}) = \emptyset$. 
	
Assume in the contrary that $\bfz \in \cD_{2,1}(\vec{x}) \cap \cD_{2,1}(\vec{y})$. Then, there exist $(a,b_1, b_2), (a',b_1', b_2')$ 
such that $$\bfz \in \cD^{(b_1,b_2) \to a}_{2,1}(\vec{x}) \cap \cD^{(b_1',b_2') \to a'}_{2,1}(\vec{y}),$$ 
and assume also that $\bfz$ is the result of deleting bits $i$ and $i+1$ from $\vec{x}$ and $j$ and $j+1$ from $\vec{y}$, and without loss of generality $i<j$. 

Since $\cC_{2,1}(n,a,c)$ is a single-deletion-correcting code, according to Claim~\ref{claim:2-1burst}, we can assume that at least one of $(a,b_1,b_2), (a', b_1', b_2')$ belongs to the set $\{(0,1,1), (1,0,0)\}$, and without loss of generality, assume that  $(a,b_1,b_2) \in \{(0,1,1), (1,0,0)\}$. First suppose $(a,b_1,b_2) =(1,0,0)$. Since $\sum_{i=1}^n x_i - \sum_{i=1}^n y_i \equiv 0~(\bmod 4)$, we have $(b_1', b_2') = (0,0)=(b_1,b_2)$. Furthermore, since $\bfz \in \cD^{(b_1,b_2) \to a}_{2,1}(\vec{x}) \cap \cD^{(b_1',b_2') \to a'}_{2,1}(\vec{y})$, $a' + b_1 + b_2 \equiv a + b_1' + b_2'~(\bmod 4)$ and so $a' = a=1$. Next, suppose $(a,b_1,b_2) =(0,1,1)$. Then, using idential logic $(b_1', b_2') = (b_1,b_2)=(1,1)$ and $a'=a=0$ so that we conclude that if one of $(a,b_1,b_2), (a', b_1', b_2')$ is in the set $\{(0,1,1), (1,0,0)\}$, then $(a,b_1,b_2)=(a', b_1', b_2')$. 

	
We consider the case where $(a,b_1,b_2)=(0,1,1)$. In this case, $\vec{x}, \vec{y}$ will have the following structure:\\ \\
	\begin{tabular}{ l l l l l l}
		$\vec{x} =$ &\hspace{-3ex} $(x_1, \ldots, x_{i-1},$ &\hspace{-2ex} $1,1,$ & \hspace{-1ex}$x_{i+2}, \ldots,x_{j},$ &\hspace{-2ex} $0,$ &\hspace{-2ex} $x_{j+2}, \ldots x_n)$,\\
		$\vec{y} =$ &\hspace{-3ex} $(y_1, \ldots, y_{i-1},$ &\hspace{-2ex} $0,$ &\hspace{-2ex} $y_{i+1}, \ldots,y_{j-1},$ &\hspace{-2ex} $1,1,$ &\hspace{-2ex} $y_{j+2}, \ldots y_n),$ \\ \\
	\end{tabular} \\
	where $x_\ell=y_\ell$ for $1\leq \ell \leq i-1$ and $j+2\leq \ell \leq n$, and $x_{i+2} = y_{i+1},$ $x_{i+3} = y_{i+2},$  $x_{i+4} = y_{i+3}, \ldots, $ $x_j= y_{j-1}$. Since $\vec{x} \neq \vec{y}$ and $j-i >0$, we have
	\begin{align*}
		& \sum_{\ell=1}^n \ell\cdot y_\ell - \sum_{\ell=1}^n \ell \cdot x_\ell = \sum_{\ell=i}^{j+1} \ell\cdot y_\ell - \sum_{\ell=i}^{j+1} \ell \cdot x_\ell  \\
		= & (2j + 1) - (2i +1) - \wt((x_{i+2}, \ldots, x_j))\\
		= & 2(j-i) -  \wt((x_{i+2}, \ldots, x_j)),
	\end{align*}
	where $\wt((x_{i+2}, \ldots, x_j))$ denotes the Hamming weight of $(x_{i+2}, \ldots, x_j)$. Since $0 \leq \wt((x_{i+2}, \ldots, x_j)) \leq j-i-1$, we conclude that
	$$2\leq j-i+1 \leq  \sum_{\ell=1}^n \ell\cdot y_\ell - \sum_{\ell=1}^n \ell \cdot x_\ell \leq 2(j-i)\leq 2(n-1),$$
	in contradiction to $\sum_{\ell=1}^n \ell\cdot y_\ell - \sum_{\ell=1}^n \ell \cdot x_\ell \equiv 0~(\bmod (2n-1))$. 
	The case where $(a,b_1,b_2)=(1,0,0)$ can be proven in a similar manner and so the details are omitted. Therefore, we conclude that 
	$\cD_{2,1}(\vec{x}) \cap \cD_{2,1}(\vec{y}) = \emptyset$ and thus $\cC_{2,1}(n,a,c)$ is a single-deletion-correcting code.
\end{IEEEproof}

The following corollary summarizes this discussion.
\begin{corollary}
For all $n\geq 4$ there exist $a \in \mathbb{Z}_{2n-1}$ and $c \in \mathbb{Z}_4$ such that the redundancy of the code $\cC_{2,1}(n,a,c)$ from Construction~\ref{constr:2-1-Code} is at most $\log(4(2n-1)) < \log(n) + 3$.
\end{corollary}

\subsection{Correcting a Burst of Length at most $b$}
We are now ready to show our constructions for $b=3,4$.
\begin{construction}\label{constr:3-non-cons-burst}
Let $\mathcal{C}_3$ denote the code from Construction~\ref{const:cons_burst} for $b=3$.
For integers $n$ and $a_1 \in \mathbb{Z}_n$, $a_2, a_3 \in \mathbb{Z}_{n-1}$, $c_2,c_3 \in \mathbb{Z}_4$, let $\mathcal{C}_{b \leq 3}(n,a_1,a_2,a_3,c_2,c_3)$ be the following code:
\begin{align*}
\mathcal{C}_{b \leq 3} \triangleq   \Big \{ \vec{x} \in  \F_2^n : \ & \vec{x} \in VT_{a_1}(n),\\
&\vec{x} \in \mathcal{C}_3, \\
& A_2(\vec{x})_1 \in \mathcal{C}_{2,1}(\frac{n}{2},a_2,c_2),\\
&A_2(\vec{x})_2 \in \mathcal{C}_{2,1}(\frac{n}{2},a_3,c_3)\Big \}.
\end{align*}
\end{construction}
\begin{theorem}\label{thm:non-cons-b=3}
The code from Construction~\ref{constr:3-non-cons-burst} can correct a non-consecutive deletion burst of size at most three.
\end{theorem}
\begin{IEEEproof}
Let $\vec{x}$ be the transmitted codeword and $\vec{y}$ is the received vector. From the length of the received vector $\vec{y}$, we know the number of deletions that occurred, denoted by $a$.
If $a=1$, the deletion can be corrected since $\vec{x}$ is a codeword of the VT-code $VT_{a_1}(n)$. If $a=3$, we have a \emph{consecutive} deletion burst of size three which can be corrected since $\vec{x}$ is a codeword in $\cC_3$, which is a three-burst-deletion-correcting code. 

If $a=2$, then the $(2,1)$-burst correcting code succeeds in any case as will be shown in the following.
If the two deletions occur consecutively, each of the two rows of the array $A_2(\vec{y})$ corresponds to a codeword from a code $\mathcal{C}_{2,1}$ with a single deletion which can be corrected.
If the two deletions occur at positions $i$ and $i+2$ (they have to be within three bits), then:
\begin{equation*}
\vec{y} = (x_1,\dots,x_{i-1}, x_{i+1},x_{i+3}, \dots, x_n)
\end{equation*}
and (assuming w.l.o.g. that $i$ is even)
\begin{equation*}
A_2(\vec{y}) = 
\begin{bmatrix}
x_1 & x_3 &\dots& x_{i-3} & x_{i-1} & x_{i+3} & \dots & x_{n-1}\\
x_2 & x_4 & \dots & x_{i-2} & x_{i+1} & x_{i+4} & \dots & x_n
\end{bmatrix}.
\end{equation*}
Compared to $A_2(\vec{x})$, the first row suffers from a single deletion ($x_{i+1}$) and the second from two deletions ($x_{i}$ and $x_{i+2}$) immediately followed by an insertion ($x_{i+1}$). This can also be corrected by the code $\mathcal{C}_{2,1}$. If $i$ is odd, there is a single deletion in the second row and two deletions followed by one insertion in the first row.
\end{IEEEproof}

\begin{theorem}
There exists a code by Construction~\ref{constr:3-non-cons-burst} which can correct a non-consecutive burst of size at most 3 with redundancy at most $ 4 \log(n) + 2\log(\log(n)) + 6$.
\end{theorem}
\begin{IEEEproof}
The set of $n+1$ VT-codes $VT_{a_1}(n)$ for $0\leq a_1\leq n$ as well as the set of $n$ codes $\mathcal{C}_{2,1}(n,a_2,c)$ and $\mathcal{C}_{2,1}(n,a_3,c)$ for $0\leq a_2,a_3\leq n-1, 0\leq c\leq 3$ form partitions of the space; i.e., $\cup_{a_1=0}^n VT_{a_1}(n) = \FTwo^n$, $\cup_{a_2=0}^{n-1} \cup_{c=0}^3 \mathcal{C}_{2,1}(n,a_2,c) = \FTwo^n$ and $\cup_{a_3=0}^{n-1} \cup_{c=0}^3 \mathcal{C}_{2,1}(n,a_3,c) = \FTwo^n$. In particular, they also form a partition of the code $\cC_3$ from Construction~\ref{const:cons_burst}. 
Therefore, by the pigeonhole principle, there are choices for $a_1,a_2,a_3,c$ such that the intersection of the three codes requires redundancy at most the sum of the redundancies of the three codes. 
\end{IEEEproof}

We now turn to the case of $b=4$, which follows the same ideas as for $b=3$, so we explain its main ideas.
\begin{construction}\label{constr:4-non-cons-burst}
Let $\mathcal{C}_4$ denote the code from Construction~\ref{const:cons_burst} for $b=4$.
For integers $n$ and $a_1,a_2 \in \mathbb{Z}_{n-1}$, $b_1,b_2,b_3 \in \mathbb{Z}_{2n/3-1}$, $c_1, c_2,d_1,d_2,d_3 \in \mathbb{Z}_4$, let $\mathcal{C}_{b \leq 4}$ be as follows:
\begin{align*}
\mathcal{C}_{b \leq 4} \triangleq   \Big \{ \vec{x} \in  \F_2^n : \ & \vec{x} \in VT_{a_1}(n),\\
&\vec{x} \in \mathcal{C}_4,\\
& A_2(\vec{x})_i \in \mathcal{C}_{2,1}(\frac{n}{2},a_i,c_i), i=1,2,\\
& A_3(\vec{x})_i \in \mathcal{C}_{2,1}(\frac{n}{3},b_i,d_i), i=1,2,3 \Big \}.
\end{align*}
\end{construction}
\begin{theorem}
The code from Construction~\ref{constr:4-non-cons-burst} can correct a non-consecutive deletion burst of size at most four.
\end{theorem}
\begin{IEEEproof}
Let $\vec{x}$ be the transmitted codeword and $\vec{y}$ is the received vector.
As for $b \leq 3$, we know the number of deletions that occurred, denoted by $a$.
If $a=1$, the deletion can be corrected since each codeword is from a VT-code. If $a=4$, we have a \emph{consecutive} deletion burst of size four which can be corrected since each codeword of $\mathcal{C}_{b \leq 4}$ is a codeword of $\mathcal{C}_4$. 
If $a=2$, the following cases can happen:
\begin{itemize}
\item The two deletions occur consecutively, then each row of $A_2(\vec{x})$ is affected by a single deletion.
\item The two deletions occur with one position in between, then one row is affected by a single deletion and the other one by a $(2,1)$-burst (similar to the proof of Theorem~\ref{thm:non-cons-b=3}).
\item There are two positions between the two deletions, i.e., positions $i$ and $i+3$ are deleted. Then:
\begin{equation*}
\vec{y} = (x_1,\dots,x_{i-1}, x_{i+1},x_{i+2},x_{i+4}, \dots, x_n)
\end{equation*}
and (assuming w.l.o.g. that $i$ is even)
\begin{equation*}
A_2(\vec{y}) = 
\begin{bmatrix}
x_1  &\dots & x_{i-1} & x_{i+2} & x_{i+5} &\dots & x_{n-1}\\
x_2  & \dots  & x_{i+1} & x_{i+4} & x_{i+6}&\dots & x_n
\end{bmatrix}
\end{equation*}
and both rows are affected by a $(2,1)$-burst.
\end{itemize}
Since the rows of $A_2(\vec{x})$ are codewords of $\mathcal{C}_{2,1}$, we can correct the deletions in any of these cases.

Similarly, for $a=3$, the following cases can happen:
\begin{itemize}
\item The three deletions occur consecutively, then each row of $A_3(\vec{x})$ is affected by a single deletion.
\item The deletions occur at positions $i$, $i+1$ and $i+3$. Then:
\begin{equation*}
\vec{y} = (x_1,\dots,x_{i-1}, x_{i+2},x_{i+4}, \dots, x_n)
\end{equation*}
and (assuming w.l.o.g. that $i$ is divisible by three)
\begin{equation*}
A_2(\vec{y}) = 
\begin{bmatrix}
x_1  &\dots& x_{i-2} & x_{i+4} & \dots & x_{n-2}\\
x_2  & \dots & x_{i-1} & x_{i+5} & \dots & x_{n-1}\\
x_2  & \dots & x_{i+2} & x_{i+6} & \dots & x_{n}\\
\end{bmatrix},
\end{equation*}
then the last row is affected by a $(2,1)$-burst and the other ones by a single deletion.
\item The deletions occur at positions $i$, $i+2$ and $i+3$. Then, similarly to before, two rows are affected by a single deletion and one row by a $(2,1)$-burst.
\end{itemize}
Since the rows of $A_3(\vec{x})$ are codewords of $\mathcal{C}_{2,1}$, we can correct the deletions in either of these cases.
\end{IEEEproof}

The next theorem summarizes this construction and its redundancy.
The redundancy follows as in Theorem~\ref{thm:non-cons-b=3} by the pigeonhole principle.
\begin{theorem}
There exists a code constructed by Construction~\ref{constr:4-non-cons-burst} with redundancy at most
$ 7 \log (n) +  2\log(\log(n)) + 4 $.

\end{theorem}

We note that for $b >4$ we cannot extend this idea and it remains as an open problem to construct efficient codes for correcting a non-consecutive burst of deletions of size $b>4$. These constructions give some first ideas to correct a burst of non-consecutive deletions/insertions. To evaluate the constructions in this section, we would like to compare the achieved redundancy with the one from~\cite{BrakensiekGuruswamiZbarksy-DeletionCorrection} which corrects arbitrary number of deletions and in particular any kind of burst. However, the paper~\cite{BrakensiekGuruswamiZbarksy-DeletionCorrection} uses asymptotic considerations which do not explicitly state the exact redundancy. Moreover, we believe that our constructions for $b\leq 4$ are more practical. 

\section{Conclusion and Open Problems}\label{sec:conc}
In this paper, we have studied codes for correcting a burst of deletions or insertions in three models. Our main contribution is the construction of binary $b$-burst-deletion-correcting codes with redundancy at most $\log(n) + (b-1)\log(\log(n)) +b -\log(b)$ bits and a non-asymptotic upper bound on the cardinality of such codes. We have extended this construction to codes which correct a consecutive burst of size at most $b$, and studied codes which correct a burst of size at most $b$ (not necessarily consecutive) for the cases $b=3,4$. While the results in the paper provide a significant contribution in the area of codes for insertions and deletions, there are still several interesting problems which are left open. Some of them are summarized as follows:
\begin{enumerate}
\item Close on the lower and upper bound on the redundancy of $b$-burst-deletion-correcting codes.
\item Constructions of better codes which correct a consecutive burst of deletion of size at most $b$.
\item Construction of codes which correct a non-consecutive deletion burst of size at most $b$, for arbitrary $b$. The best codes are the ones which correct any $b$ deletions from~\cite{BrakensiekGuruswamiZbarksy-DeletionCorrection}.
\item Find better lower bounds on the redundancy of codes which correct a burst of deletions in the two last models (the only lower bound is the one for $b$-burst-deletion-correcting codes).
\item Generalize all our constructions to more than one burst of deletions or insertions.
\end{enumerate}

\appendices
\section{Calculating the value of $N(n,b,i)$}\label{app:N(n,bmi)}
In this appendix we calculate the value of $N(n,b,i) = |\{ \vec{x}\in \FTwo^{n} : |D_b(\vec{x})| =i \}|$. 
\begin{lemma}\label{lem:vectors-same-burst-ball-size}
For $1\leq i\leq n-b+1$ we have that 
\begin{equation*}
N(n,b,i) = 2^b \binom{n-b}{i-1}.
\end{equation*}
\end{lemma}
\begin{IEEEproof}
Recall that we can arrange a vector $\vec{x} = (x_1,x_2,\dots,x_n)$ into a $b \times \frac{n}{b}$ array $ A_b(\vec{x}) $.

Let $ r(\textbf{x}_j) $ denote the number of runs in the $j$th row of $ A_b(\vec{x})$. From equation (\ref{eq:ball_size}), we have that
\[ |D_b(\textbf{x})|= \left(\sum_{j=1}^{b}r(\textbf{x}_j)\right) - b + 1.  \]
Thus, counting the number of vectors of length $n$ whose $b$-burst deletions ball size is $i$ is equivalent to counting the number of vectors of length $n$ for which  \[ \left(\sum_{j=1}^{b}r(\textbf{x}_j)\right) = i+b-1 . \]
The number of binary vectors of length $ n $ with $ r $ runs is  \[ 2\binom{n-1}{r-1} \triangleq M(n,r). \]
For $ b=2 $, $ N(n,2,i) $ is given by
\begin{align*}
&\sum_{0<r_1,r_2: r_1+r_2=i+2-1}M\left(\frac{n}{2},r_1\right)\cdot M\left(\frac{n}{2},r_2\right)\\
&=\sum_{r_1=1}^{i}M\left(\frac{n}{2},r_1\right)\cdot M\left(\frac{n}{2},i+1-r_1\right)\\
&=\sum_{r_1=1}^{i}2\binom{\frac{n}{2}-1}{r_1-1}\cdot 2\binom{\frac{n}{2}-1}{i-r_1}\\
&=4\sum_{r_1=0}^{i-1}\binom{\frac{n}{2}-1}{r_1}\cdot\binom{\frac{n}{2}-1}{i-1-r_1}\\
&= 4\binom{n-2}{i-1}.
\end{align*}
We used Vandermonde's identity in the final step which states that for any nonnegative integer $ n $ the following relation holds true:
\[ \sum_{k=0}^{n}\binom{x}{k}\binom{y}{n-k}=\binom{x+y}{n}. \]

We prove lemma's statement by induction on $ b $. We have already established the base case for $ b=2 $ (the $ b=1 $ case is trivially given by $ M(n,r) $).

Assume the following holds for $ b=k $:
\begin{align*}
&\sum_{\substack{0<r_1,r_2,\ldots,r_k: \\ r_1+r_2+\ldots+r_k=i+k-1}} M\left( \frac{n}{k},r_1 \right)\cdot M\left( \frac{n}{k},r_2 \right) \cdots M\left( \frac{n}{k},r_k \right)\\
&= 2^k \binom{n-k}{i-1}.
\end{align*}   
We wish to show that for $ b=k+1 $,
\begin{align*}
&\sum_{\substack{0<r_1,r_2,\ldots,r_{k+1}: \\ r_1+r_2+\ldots+r_{k+1}=i+k}} M\left( \frac{n}{k+1},r_1 \right)\cdot M\left( \frac{n}{k+1},r_2 \right) \\
&\cdots M\left( \frac{n}{k+1},r_{k+1} \right)= 2^{k+1} \binom{n-(k+1)}{i-1}.
\end{align*}

Let us now prove the previous equation using the inductive assumption:
\begin{align}
&\sum_{\substack{0<r_1,r_2,\ldots,r_{k+1}:\\ r_1+r_2+\ldots+r_{k+1}=i+k}} M\left( \frac{n}{k+1},r_1 \right)\cdot M\left( \frac{n}{k+1},r_2 \right) \nonumber\\
&\cdots M\left( \frac{n}{k+1},r_{k+1} \right)\nonumber\\
&=\sum_{r_{k+1}=1}^{i}M\left( \frac{n}{k+1},r_{k+1} \right)\nonumber\\
&\cdot \sum_{\substack{0<r_1,r_2,\ldots,r_k: \\ r_1+r_2+\ldots+r_k=i+k-r_{k+1}}} M\left( \frac{n}{k+1},r_1 \right) \cdots M\left( \frac{n}{k+1},r_k \right)\label{eqn2}\\
&=\sum_{r_{k+1}=1}^{i}M\left( \frac{n}{k+1},r_{k+1} \right) \cdot 2^k \binom{\frac{nk}{k+1}-k}{i-r_{k+1}}\label{eqn3}\\
&=\sum_{r_{k+1}=1}^{i} 2 \binom{\frac{n}{k+1}-1}{r_{k+1}-1} \cdot 2^k \binom{\frac{nk}{k+1}-k}{i-r_{k+1}}\nonumber\\
&=2^{k+1}\sum_{r_{k+1}=0}^{i-1}\binom{\frac{n}{k+1}-1}{r_{k+1}} \cdot \binom{\frac{nk}{k+1}-k}{i-r_{k+1}-1}\nonumber\\
&=2^{k+1}\binom{\frac{n}{k+1}-1+\frac{nk}{k+1}-k}{i-1}\nonumber\\
&=2^{k+1}\binom{n-(k+1)}{i-1}.\nonumber
\end{align}
We used the induction assumption to simplify~\eqref{eqn2} to~\eqref{eqn3}.

%
%
%
\end{IEEEproof}

\section{Encoding of Run-Length-Limited Sequences}\label{RLL_encoding}
In this appendix we describe how to efficiently encode vectors that satisfy the $(\log(n)+3)$-RLL$(n)$ constraint. Namely, Algorithm~\ref{alg:encoding} uses one redundancy bits in order to encode vectors of maximum run length at most $\lceil\log(n)\rceil+3$.

\begin{algorithm}
	\caption{Run-Length Encoding}\label{alg:encoding}
	\begin{algorithmic}[1]
		\vspace{.1ex}
		\Require Sequence $\vec{x}\in \FTwo^n$ 
		\Ensure Sequence $\vec{y} \in \FTwo^{n+1}$ with run length $\leq \lceil\log(n)\rceil+3$
		\State Define $\vec{y} = (x_1,x_2,\dots,x_{n}, {0}) \in \FTwo^{n+2}$
		\State Set $i=1$ and $i_{end} = n$
		\While{$i \leq i_{end}$}
		\If{length of run starting at $y_i$ is $\geq \lceil\log(n)\rceil\!+\!4$}
		\State $p(i)$: binary representation of $i$ with $\lceil\log(n)\rceil$ bits
		\State remove $\lceil\log(n)\rceil+3$ bits of this run from $\vec{y}$
		\State append $(1,p(i),01)$ on the right of $\vec{y}$
		\State set $i_{end}=i_{end}-\log(n)-3$
		\Else 
		\State set $i=i+1$
		\EndIf
		\EndWhile
	\end{algorithmic}
\end{algorithm}
Notice that in Algorithm~\ref{alg:encoding} if there is a run of length at least $a \cdot (\lceil\log(n)\rceil+3)+1$, for some $a \geq 2$, then the same vector $(1,p(i),01)$ is appended $a$ times. 

\begin{theorem}\label{thm:encoding}
	Given any sequence $\vec{x} \in \FTwo^n$, Algorithm~\ref{alg:encoding} outputs a sequence $\vec{y}\in \FTwo^{n+1}$ where any run has length at most $\lceil\log(n)\rceil+3$ and such that $\vec{x}$ can uniquely be reconstructed given $\vec{y}$.
\end{theorem}
\begin{IEEEproof}
	First, let us explain the length of $\vec{y}$.
	Some runs of length $\lceil\log(n)\rceil+3$ are removed and a block $(1,p(i),01)$ is appended. Both blocks have length $\lceil\log(n)\rceil+3$, so this does not change the length of the vector and we have only one additional bit, which is the zero bit that was appended in Step~1.
	
	Second, let us consider the maximum run length.
	The longest run in $\vec{y}$ is of length $\lceil\log(n)\rceil+3$, since any longer run is removed and replaced by $(1,p(i),01)$. Clearly, in the newly appended blocks, the run length is at most $\lceil\log(n)\rceil+1$ due to the "$01$". The first "$1$" in $(1,p(i),01)$ is necessary to avoid the following case: the sequence $\vec{x}$ ends with $\log(n)$ zeros and there is a sequence of $2\log(n)$ zeros at the beginning. We have to write the number zero in binary to the right of the redundancy bit. This would create a sequence of $2\log(n)+1$ zeros if the first one of $(1,p(i),01)$ was not there.
	
	To reconstruct $\vec{x}$ given $\vec{y}$, we start from the right. Check if the rightmost bit is $0$ or $1$. 
	If it is $0$, then the leftmost $n$ bits of $\vec{y}$ are equal to $\vec{x}$. 
	If it is $1$, we know that the rightmost $\lceil\log(n)\rceil+3$ bits are an encoded block, where $p(i)$ provides the position where to insert a run of length $\lceil\log(n)\rceil+3$.
	The value of this run is the value of the bit at position $i$. We can therefore insert such a run and remove the rightmost $\lceil\log(n)\rceil+3$ bits. Then, we check again the rightmost bit. We repeat the previous strategy until the rightmost bit is $0$, in which case the first $n$ bits correspond to $\vec{x}$ we and have decoded our original sequence.
\end{IEEEproof}

\begin{example}
	Let $n=16$ and therefore $\log(n) = 4$ and $\log(n)+3 = 7$. Consider the following sequence:
	\begin{equation*}
	\vec{x} = (0111 1111 1111 1111),
	\end{equation*}
	where the middle one-run has length $15$. Let us go through the steps of Algorithm~\ref{alg:encoding}.
	\begin{enumerate}
		\item $\vec{y} = (0111 1111 1111 1111 {0})$
		\item $i=1$ and $i_{end} = 16$.
		\item for $i=1$: do nothing.
		\item $i=2$: the run starting at $x_2$ is at least $8$ bits long.\\ 
		Define $p(2) = (0010)$, remove $7$ bits from the one run in $\vec{y}$ and append $(1001001)$.\\
		Thus, $\vec{y} = (0 1111 1111 {0} 1001001)$. \\
		$i_{end} = 16-7=9$.
		\item $i=2$: the run starting at $x_2$ is $8$ bits long.\\
		Define $p(2) = (0010)$, remove $7$ bits from the one run in $\vec{y}$ and append $(1001001)$.\\
		Thus, $\vec{y} = (0 1 {0} 10010011 10010011)$.\\
		$i_{end} = 9-7=2$.
		\item $i=2$: do nothing and then the while-loop stops.
	\end{enumerate}
	The decoding works as described in the proof of Theorem~\ref{thm:encoding}.
\end{example}

\section{Decoder of Shifted VT Codes}\label{SVT_decoder}
In order to better understand the rationale behind the SVT-code, let us explore the details of the decoding algorithm (presented in pseudocode form in Algorithm 1).

\begin{algorithm}
\caption{Decoding algorithm for the $ SVT_a(n,P) $ code}\label{alg:euclid}
\begin{algorithmic}[1]
\Require Received vector \textbf{y}, integers \textit{a, u, P}
\Ensure Corrected vector \textbf{y} (equal to original vector \textbf{x})
\State $DelVal\gets wt(\textbf{y})~(\bmod 2)$
\State $ \hat{\textbf{y}} \gets (y_u, y_{u+1},\ldots,y_{u+P-2}) $
\State $ a' \gets \sum\limits_{i=1}^{u+P-2} iy_i + \sum\limits_{i=u+P-1}^{n-1} (i+1)y_i~(\bmod P) $
\State $ \Delta \gets a-a'~(\bmod P) $
\If{$ DelVal=0 $}
\State $ DelPos \gets $ first position to the left of $ \Delta $ 1's in $ \hat{\textbf{y}} $
\Else 
\State $ DelPos \gets $ first position to the right of $ \Delta-u-wt(\hat{\textbf{y}})~(\bmod P) $ 0's in $ \hat{\textbf{y}} $
\EndIf
\State Insert $ DelVal $ into position $ DelPos $ of $ \hat{\textbf{y}} $
\end{algorithmic}
\end{algorithm}	

The decoder receives the vector $ \textbf{y}=(y_1,\ldots,y_{n-1}) \in \FTwo^{n-1} $ which is the vector \textbf{x} with a single bit deleted. The decoder knows the first possible location of the deleted bit, \textit{u}, as well as the number of possible positions of the deleted bit, \textit{P}. In our overall code construction, the parameter \textit{a}, the weighted sum from Definition 1, and \textit{P} are both known to the decoder ahead of time, while \textit{u} is gleaned from decoding the first row of our codeword array. The value of the deleted bit, \textit{DelVal}, is found by simply checking the overall parity of the received vector.

We define $ \hat{\textbf{y}} = (y_u, y_{u+1},\ldots,y_{u+P-2}) $. This vector contains the $ P-1 $ bits in which we are not certain about their position in \textbf{x}. Any bit in position $ i,i<u $ are in their proper positions, and any bit in position $ i, i>u+P-2 $ will be shifted one position to the right once we insert the deleted bit.

In the decoding algorithm, $ a' $ is the \textit{augmented} weighted sum of our received vector \textbf{y}. We define the difference between the original weighted sum of \textbf{x} and our augmented weighted sum of \textbf{y} as $ \Delta $. Since our calculation of $ a' $ properly weighted every bit outside of $ \hat{\textbf{y}} $, we can focus our attention solely on $ \hat{\textbf{y}} $, i.e., inserting a bit to increase the weighted sum of $ \hat{\textbf{y}} $ by $ \Delta $ also increases the weighted sum of \textbf{y} by $ \Delta $ (thus yielding \textbf{x}).

Within $ \hat{\textbf{y}} $, let us denote the number of 0's and 1's to the left of the bit we insert as $ L_0 $ and $ L_1 $, respectively. Similarly, let us call the number of 0's and 1's to the right of the bit we insert as $ R_0 $ and $ R_1 $.

Inserting a 0 into $ \hat{\textbf{y}} $ increases its weighted sum by $ R_1~(\bmod P) $ since all the 1's are shifted one space to the right. Note that this is true even if the 1 is pushed from weight $ P-1 $ to weight $ P~(\bmod P) =0 $.
Thus, if a 0 was deleted, we insert a 0 in the first space to the left of $ \Delta $ 1's.

Inserting a 1 into the $ \textit{i}$th position of $ \hat{\textbf{y}} $ increases its weighted sum by $ R_1+i+u-1~(\bmod P) $. Since $ i=L_0+L_1+1 $, this implies $ \Delta=R_1+L_1+L_0+u \bmod P $. Since $ wt(\hat{\textbf{y}}) = L_1+R_1 $, we have $ \Delta=L_0+wt(\hat{y})+u~(\bmod P) $. Solving for $ L_0 $ yields $ L_0 = \Delta - u - wt(\hat{\textbf{y}})~(\bmod P)$. Thus, if the deleted bit was a 1, we insert a 1 in the first space to the right of $ \Delta - u - wt(\hat{\textbf{y}})~(\bmod P)$ 0's in $ \hat{\textbf{y}} $.

In the following example, the transmitted vector \textbf{x} is encoded as an $ SVT_0(16) $ codeword. Additionally, let us assume that the first row of our codeword array was encoded to have the longest run be no greater than 4, thus we have $ P=5 $. Also, let us assume that after correcting the first row, we find $ u=8 $. Note that the following is an example of decoding any row in our codeword array besides the first row.
\begin{example}
Let us assume the transmitted vector was the following $ SVT_0(16) $ codeword: $ \textbf{x}=(11110110\textbf{0}1100011) $. Based on previous information, the decoder knows $ P=5 $ and $ u=8 $. During transmission, the 9th bit was deleted (bolded), so the received vector was $ \textbf{y}=(1111011\underline{0110}0011) $. The receiver determines the value of the deleted bit:
\[  DelVal = wt(\textbf{y})~(\bmod 2) = 10~(\bmod 2) = 0 . \] 
The receiver calculates the augmented weighted sum of the received vecor $ a'=3 $. Now the receiver calculates the differences in the weighted sums:
\[  \Delta = a-a' ~(\bmod 5) = 0-3~(\bmod 5) = 2 . \]
Since $ u=8 $, we have $ \hat{\textbf{y}}=(0110) $, underlined in \textbf{y}. Since $ DelVal=0 $, $ DelPos $ is the first position to the left of $ \Delta=2 $ 1's in $ \hat{\textbf{y}} $, yielding $ \hat{\textbf{y}}=(0\textbf{0}110) $. With the insertion of this bit, we have successfully decoded the original sent codeword \textbf{x}.
\end{example}

\section*{Acknowledgement}
C. Schoeny's work was funded in part by the NISE program at SSC Pacific.

A. Wachter-Zeh was supported by the European Union’s Horizon 2020 research and innovation programme under the Marie Sklodowska-Curie grant agreement No. 655109. 

E. Yaakobi's work was supported in part by the Israel Science Foundation (ISF) grant No. 1624/14.

R. Gabrys' work was funded in part by the NISE program at SSC Pacific.

\bibliographystyle{IEEEtranS}
\bibliography{mybib}
\end{document}

%% file: mydefs.tex
\newtheorem{theorem}{Theorem}
\newtheorem{definition}{Definition}
\newtheorem{lemma}{Lemma}
\newtheorem{corollary}{Corollary}

\newtheorem{remark}{Remark}

\newtheorem{example}{Example}

\newtheorem{construction}{Construction}
\newtheorem{claim}{Claim}

 \newcommand{\qed}{\hfill \mbox{\raggedright \rule{.07in}{.1in}}}

\newcommand{\Fq}{\ensuremath{\mathbb F_{q}}}
\newcommand{\FTwo}{\ensuremath{\mathbb F_{2}}}

\newcommand{\F}{\ensuremath{\mathbb F}}

\newcommand{\myset}[1]{\mathcal{#1}}

\renewcommand{\bar}{\overline}

\DeclareMathOperator{\wt}{wt}

\renewcommand{\vec}[1]{\ensuremath{\textbf{#1}}}

\newcommand{\mycode}[1]{\ensuremath{\mathcal{#1}}}

\newcommand{\bfa}{{\boldsymbol a}}

\newcommand{\bfz}{{\boldsymbol z}}

\newcommand{\cC}{{\cal C}}
\newcommand{\cD}{{\cal D}}

%% file: burst-deletions_Journal_Final_Single_Col.bbl
\begin{thebibliography}{10}
\providecommand{\url}[1]{#1}
\csname url@samestyle\endcsname
\providecommand{\newblock}{\relax}
\providecommand{\bibinfo}[2]{#2}
\providecommand{\BIBentrySTDinterwordspacing}{\spaceskip=0pt\relax}
\providecommand{\BIBentryALTinterwordstretchfactor}{4}
\providecommand{\BIBentryALTinterwordspacing}{\spaceskip=\fontdimen2\font plus
\BIBentryALTinterwordstretchfactor\fontdimen3\font minus
  \fontdimen4\font\relax}
\providecommand{\BIBforeignlanguage}[2]{{%
\expandafter\ifx\csname l@#1\endcsname\relax
\typeout{** WARNING: IEEEtranS.bst: No hyphenation pattern has been}%
\typeout{** loaded for the language `#1'. Using the pattern for}%
\typeout{** the default language instead.}%
\else
\language=\csname l@#1\endcsname
\fi
#2}}
\providecommand{\BIBdecl}{\relax}
\BIBdecl

\bibitem{Bours-InsertionsDeletions}
P.~A. Bours, ``Codes for correcting insertions and deletion errors,'' PhD
  thesis, Eindhoven University of Technology, Jun. 1994.

\bibitem{BrakensiekGuruswamiZbarksy-DeletionCorrection}
\BIBentryALTinterwordspacing
J.~Brakensiek, V.~Guruswami, and S.~Zbarsky, ``Efficient low-redundancy codes
  for correcting multiple deletions,'' \emph{CoRR}, vol. abs/1507.06175, 2015.
  [Online]. Available: \url{http://arxiv.org/abs/1507.06175}
\BIBentrySTDinterwordspacing

\bibitem{ChengSwartFerreiraAbdelGhaffar-ThreeOrMoreBurstDelIns}
L.~Cheng, T.~G. Swart, H.~C. Ferreira, and K.~A.~S. Abdel-Ghaffar, ``{Codes for
  correcting three or more adjacent deletions or insertions},'' in \emph{Proc.
  IEEE Int. Symp. Inf. Theory (ISIT)}, Jun. 2014, pp. 1246--1250.

\bibitem{cullina2012coloring}
D.~Cullina, A.~A. Kulkarni, and N.~Kiyavash, ``A coloring approach to
  constructing deletion correcting codes from constant weight subgraphs,'' in
  \emph{Proc. IEEE Int. Symp. Inf. Theory (ISIT)}, Jul. 2012, pp. 513--517.

\bibitem{dandashi2007tactical}
F.~Dandashi, A.~Griggs, J.~Higginson, J.~Hughes, W.~Narvaez, M.~Sabbouh,
  S.~Semy, and B.~Yost, ``Tactical edge characterization framework,''
  \emph{MITRE Technical Report MTR070331}, 2007.

\bibitem{Immink91}
K.~Immink, \emph{{Coding techniques for digital recorders}}.\hskip 1em plus
  0.5em minus 0.4em\relax Prentice Hall, College Div., 1991.

\bibitem{jeong2003forward}
J.~Jeong and C.~T. Ee, ``Forward error correction in sensor networks,''
  \emph{University of California at Berkeley}, 2003.

\bibitem{KulkarniKiyavash-UpperBoundsForDeletion}
A.~A. Kulkarni and N.~Kiyavash, ``{Nonasymptotic Upper Bounds for Deletion
  Correcting Codes},'' \emph{IEEE Trans. Inf. Theory}, vol.~59, no.~8, pp.
  5115--5130, Aug. 2013.

\bibitem{Levenshtein-binarycodesCorrectingDeletions}
V.~Levenshtein, ``Binary codes capable of correcting deletions, insertions and
  reversals (in russian),'' \emph{Doklady Akademii Nauk SSR}, vol. 163, no.~4,
  pp. 845--848, 1965.

\bibitem{Levenshtein-TwoAdjacentDeletions}
------, ``Asymptotically optimum binary code with correction for losses of one
  or two adjacent bits,'' \emph{Systems Theory Research (translated from
  Problemy Kibernetiki)}, vol.~19, pp. 293--298, 1967.

\bibitem{renyi1970probability}
A.~R{\'e}nyi, \emph{{Probability Theory}}.\hskip 1em plus 0.5em minus
  0.4em\relax Budapest, Akad. Kiad\'o, 1970.

\bibitem{schilling1990longest}
M.~F. Schilling, ``{The longest run of heads},'' \emph{College Math. J},
  vol.~21, no.~3, pp. 196--207, 1990.

\bibitem{BurstDeletions-ISIT}
C.~Schoeny, A.~Wachter{-}Zeh, R.~Gabrys, and E.~Yaakobi, ``Codes for correcting
  a burst of deletions or insertions,'' in \emph{to appear Proc. IEEE Int.
  Symp. Inf. Theory (ISIT)}, Jul. 2016.

\bibitem{Sloane01onsingle-deletion-correcting}
N.~J.~A. Sloane, ``On single-deletion-correcting codes,'' in \emph{Proc. Codes
  and Designs}, 2001, pp. 273--291.

\bibitem{tenengolts1984nonbinary}
G.~Tenengolts, ``Nonbinary codes, correcting single deletion or insertion
  (corresp.),'' \emph{IEEE Transactions on Information Theory}, vol.~30, no.~5,
  pp. 766--769, 1984.

\bibitem{VarshTene-SingleDeletion1965}
R.~R. Varshamov and G.~M. Tenengolts, ``Codes which correct single asymmetric
  errors (in russian),'' \emph{Automatika i Telemkhanika}, vol. 161, no.~3, pp.
  288--292, 1965.

\end{thebibliography}
